\documentclass[10pt,twocolumn,secnumarabic,amssymb,superscriptaddress,nobibnotes,nofootinbib,aps,pra,showkeys,floatfix]{revtex4-1}

\usepackage{mathrsfs}
\usepackage{physics}
\usepackage{tikz}
\usepackage[caption=false]{subfig}
\usepackage[colorlinks=true,linkcolor=blue,citecolor=blue,urlcolor=blue]{hyperref}
\usepackage{cleveref}
\usepackage{amsmath,amssymb,amsfonts,amsthm}
\usepackage{graphicx}
\usepackage{ragged2e}
\usepackage{bm}
\usepackage{svg}
\usepackage{extarrows}
\usepackage{xcolor}
\graphicspath{{figures/}}

\usepackage{dsfont}
\usepackage[title]{appendix}

\newif\ifshowrevisions
\showrevisionstrue 

\makeatletter
\let\auto@bib\@empty
\let\auto@bib@innerbib\@empty
\makeatother

\def\rar{\rightarrow}
\def\lrar{\longrightarrow}

\def\xlrar{\xlongrightarrow}

\def\ra{\rangle}

\def\no{\nonumber}
\def\bea{\begin{eqnarray}}
\def\eea{\end{eqnarray}}
\def\be{\begin{equation}}
\def\ee{\end{equation}}

\usepackage{filecontents}

\begin{document}

\title{Work extraction from long-lived quantum coherence of a three-level system}%

\author{Wenjing Chen}%
\affiliation{School of Physics, Xi'an Jiaotong University, Xi'an, Shaanxi 710049, China}
\author{Si-Wei Han}
\affiliation{School of Physics, Xi'an Jiaotong University, Xi'an, Shaanxi 710049, China}
\author{Ziye Wang}
\affiliation{School of Physics, Xi'an Jiaotong University, Xi'an, Shaanxi 710049, China}
\author{Xiaoshan Feng}
\affiliation{School of Physics, Xi'an Jiaotong University, Xi'an, Shaanxi 710049, China}
\author{Jun Feng}
\email[Corresponding author: ]{j.feng@xjtu.edu.cn}
\affiliation{School of Physics, Xi'an Jiaotong University, Xi'an, Shaanxi 710049, China}
\affiliation{MOE Key Laboratory for Non-equilibrium Synthesis and Modulation of Condensed Matter, Xi’an 710049, Shaanxi, China}
\affiliation{Hefei National Laboratory, Hefei 230088, Anhui, China}

\begin{abstract}
We analyze work-extraction protocols using the long-lived quantum coherence of a three-level quantum system coupled to a thermal bath. We identify conditions under which persistent quantum coherence can arise, i.e., for systems with degenerate excited states and aligned transition dipoles, or for nearly degenerate systems with small energy splittings. By proposing two thermodynamic protocols that employ energy-preserving unitary operations, we show that quantum coherence can be converted into population asymmetry, thereby serving as a quantum resource for work extraction. Specifically, we propose an optimized protocol to extract the maximum extractable work (MEW), measured by the free energy difference (FED), from quantum coherence in a single-shot thermodynamic cycle. Our results highlight the thermodynamic advantages of long-lived coherence in a three-level quantum system and could influence future designs of coherence-driven quantum thermal machines.

\end{abstract}

\maketitle

\section{Introduction}

Quantum thermodynamics seeks to understand how thermodynamic laws emerge from quantum mechanics and how genuinely quantum resources may influence microscopic energy-conversion processes \cite{Vinjanampathy2016}. Among these resources, quantum coherence, manifested by off-diagonal density-matrix elements in a fixed energy eigenbasis, plays a central role in quantum information processing and quantum thermodynamics \cite{Goold2016}.

Due to inevitable interactions with the environment, quantum coherence is generally destroyed in open quantum systems. {However, an intriguing exception occurs in some multilevel configurations, where degeneracy sustains coherence via the Fano interference between indistinguishable transition pathways. A particularly enlightening model is a three-level \textsf{V}-type system, comprising a single ground state coupled by the system-bath interaction to a pair of excited states \cite{Tscherbul2014,Koyu2018,Dodin2018,Koyu2021}. When the excited states are exactly degenerate with perfectly aligned transition dipoles, the bath cannot distinguish the two transition pathways. The resulting cross-damping interference produces a bright/dark-state structure and can sustain steady \emph{noise-induced coherence} during continuous thermal contact. On the other hand, for a small but nonzero level splitting, the interference instead can generate long-lived metastable coherence over the partial-secular time window.} This dynamical effect can greatly influence the system's steady-state behavior and has been widely studied in many fields \cite{Cao2016,Avisar2016,Segal2018,Cao2018,Brumer2018,Cresser2021}, such as optics, quantum heat engines, and biological photosynthetic processes.

While most studies have emphasized the dynamical role of noise-induced coherence in specific quantum processes, its resource nature (i.e., its consumption and interconvertibility \cite{Streltsov2015,Chitambar2016,Gu2016}) has rarely been explored in a rigorous quantification framework \cite{Plenio2017}. In particular, from a quantum thermodynamic perspective, it is legitimate to ask whether and how the noise-induced quantum coherence can be converted into thermodynamic work \cite{Lostaglio2015}. {Besides its fundamental interest, this question is also intimately relevant to resource manipulation in real quantum devices, such as quantum batteries \cite{Campaioli2024,Ferraro2026}, where coherence induces limitations on their charging, capacity, and open-system energy extraction \cite{Quach2023,Yang2023,Lai2024,Oularabi2025}. For instance, while finite-time surviving quantum coherence in a two-level open system constrains the allowable operating times in experiments \cite{Li2025}, the steady-state coherence of the \textsf{V}-type system, once established, is time-independent and may offer a route to circumvent this time-window restriction, thereby providing a robust and reproducible resource for practical work extraction.}

In this paper, we address the optimal work-extraction protocols for the long-lived quantum coherence of a \textsf{V}-type system coupled to a thermal bath via a dipole-monopole interaction \cite{Tscherbul2014}. {Our interest in the three-level \textsf{V} system is twofold: (i) closed-form analytic expressions for steady noise-induced coherence (absent in two-level open systems) can be derived for the three-level \textsf{V} model \cite{Dodin2018}, enabling us to explicitly construct work-extraction cycles, and (ii) the three-level model can be regarded as a “building block” for certain $N$-level systems with $N-1$-level degeneracy \cite{Niedenzu2015}, allowing the present work-extraction protocols to be adapted to multi-level settings.} We first explore the conditions under which long-lived quantum coherence can be generated in the presence of bath noise and sustained in the steady state of the \textsf{V} system. Then, using energy-preserving unitary rotation within the degenerate excited subspace \cite{Korzekwa2016}, we propose two thermodynamic protocols that coherently manipulate population asymmetry and subsequently convert it into useful work. In particular, Protocol 1 is designed to start from the coherent asymptotic state of an exactly degenerate \textsf{V} system. Work can be extracted over repeated cycles, at the cost of consuming quantum coherence. Protocol 2 is refined to an optimized single quasistatic cycle that can fully convert the available free-energy difference (FED), including quantum coherence, into work. We show that these innovative protocols enable efficient extraction of quantum resources and demonstrate that practical thermodynamic protocols may apply to future quantum device design \cite{Deffner2022}.


The paper is organized as follows: In Sec.\ref{sec2}, we introduce the theoretical model of a three-level \textsf{V}-type open quantum system and discuss how quantum coherence is generated and sustained in a (nearly) degenerate system contacting a thermal bath. Sec.\ref{sec3} presents in detail two thermodynamic protocols for efficiently extracting work from quantum coherence. Specifically, by implementing an idealized thermodynamic cycle, we propose a single-shot protocol that converts the entire FED into practical work in a single cycle. In Sec.\ref{sec5}, we summarize our findings. For simplicity, throughout the analysis, we use the units where $\hbar=k_B=c=1$.

\section{Dynamics of Three-Level Open Quantum System}
\label{sec2}

Multilevel open quantum systems provide a versatile platform with rich dynamical properties due to their interaction with the environmental bath, a feature frequently encountered in quantum optics \cite{Scully}. A widely studied model is the \textsf{V}-type three-level atom \cite{Tscherbul2014,Dodin2018,Koyu2018,Koyu2021}, composed of two (nearly) degenerate excited states $\ket{1}$ and $\ket{2}$, and a single ground state $\ket{0}$.

\subsection{Degenerate three-level system}

We first consider a degenerate \textsf{V}-type three-level atom coupled to a thermal bath via dipole-monopole interaction. In the atom-field picture, the total Hamiltonian of the three-level system and bath can be written as
\be
H=H_S+H_\Phi+g H_I,
\label{eq.0}
\ee
where $H_\Phi$ is the free Hamiltonian of the environment field. The degenerate three-level atom Hamiltonian is
\begin{equation}
H_S = \omega \left( \ket{1}\bra{1} + \ket{2}\bra{2} \right), \label{eq.1}
\end{equation}
where $\omega$ is the energy difference between the excited and ground states. 

\begin{figure}[hbtp]
\centering  
\includegraphics[width=0.25\textwidth]{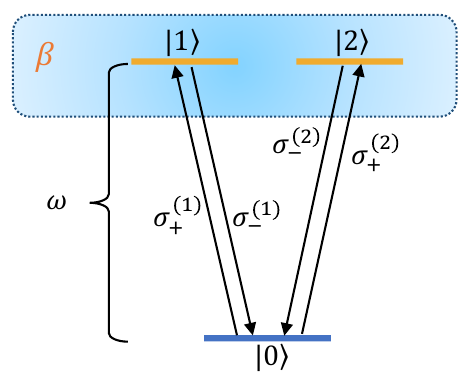}
\caption{Scheme of a degenerate three-level \textsf{V} atom with enegy spacing $\omega$. {The transition pathways from degenerate excited states ($|1\rangle$ and $|2\rangle$) to the ground state are dipole-allowed. The background field represents a thermal reservoir at inverse temperature $\beta$.}} 
\label{Degenerate 3-level}
\end{figure}

For the degenerate system, we consider the system-bath interaction that allows both transitions $\ket{1} \leftrightarrow \ket{0}$ and $\ket{2} \leftrightarrow \ket{0}$ to exchange excitations with the environment. This makes the dynamics of a three-level quantum system significantly more involved than those of the widely studied two-level qubit. The system-bath interaction is then modeled as \cite{Open1}
\begin{equation}
H_I = \sum_{i=1}^{2} \sum_{\alpha = x, y, z} \left( \sigma_{+}^{(i)} d_{(i)}^\alpha \otimes \Phi_{\alpha} + \sigma_{-}^{(i)} d_{(i)}^\alpha \otimes \Phi_{\alpha}^\dagger \right). \label{eq:HI}
\end{equation}
Here $\sigma_{+}^{(i)} = \ket{i}\bra{0}$ and $\sigma_{-}^{(i)} = \ket{0}\bra{i}$ are the transition operators, lowering or raising states between $|i\ra$ and $|0\ra$ (see Fig.\ref{Degenerate 3-level}). The vector $\mathbf{d}_{(i)} = (d_i^x, d_i^y, d_i^z)$ represents the dipole moment of the system, which tensor-products the bath operators $\Phi_{\alpha}$, corresponding to the $\alpha$-component of the environment field. {It is worth noting that for a thermal bath, $\Phi_\alpha=\Phi_\alpha^{\dagger}$, indicating that the interaction Hamiltonian \eqref{eq:HI} retains both the co-rotating and counter-rotating terms (e.g., $\sigma_{-} \Phi^{\dagger}$ and $\sigma_{+} \Phi$) at the Hamiltonian level, without employing the rotating-wave approximation (RWA).} At equilibrium with temperature $T$, the thermal field operators satisfy the Kubo-Martin-Schwinger (KMS) condition \cite{HaagHugenholtzWinnink1967,Pusz1978}:
\begin{equation}
\langle \Phi_{\alpha}(t) \Phi_{\beta}(t') \rangle = \langle \Phi_{\beta}(t') \Phi_{\alpha}(t + i \beta) \rangle,
\label{eq.3}
\end{equation}
with inverse temperature $\beta := 1 / T$.

The three-level atom with dipole-monopole interaction \eqref{eq:HI} can be treated as an open quantum system. {In the weak-coupling limit ($g \ll 1$) and for short bath correlation times ($\tau_B \ll \tau_R$) compared to the relaxation time $\tau_R \sim 1 / \gamma_{\pm}$, the Born-Markov construction is justified. In this regime, the counter-rotating terms oscillate rapidly relative to the dissipative evolution and average out under the condition $\gamma_{ \pm} \ll \omega_i$ \cite{Han2025}, enabling the employment of the secular approximation.} Consequently, the reduced dynamics of the atom density matrix $\rho$ is governed by the Gorini-Kossakowski-Sudarshan-Lindblad (GKSL) master equation \cite{sec1-15,sec1-16} as:
\begin{equation}
\frac{d\rho}{dt} = -i[H_S + H_{\text{LS}}, \rho] + \mathcal{L}[\rho],
\label{master}
\end{equation}
where $H_{\text{LS}}$ is the Lamb shift Hamiltonian accounting for energy level renormalization due to the environment, and $\mathcal{L}[\rho]$ is the dissipator that encapsulates the irreversible dynamics induced by the thermal bath.

Since two excited states are degenerate, they couple identically to the thermal reservoir, making the dissipator take the form \cite{Dodin2018}:
\begin{equation}
\begin{aligned}
\mathcal{L}_1[\rho] = & \sum_{i,j=1}^{2} \Gamma_{(ij)}^+ \left( \sigma_{-}^{(i)} \rho \sigma_{+}^{(j)} - \frac{1}{2} \{ \sigma_{+}^{(j)} \sigma_{-}^{(i)}, \rho \} \right) \\
& + \Gamma_{(ij)}^- \left( \sigma_{+}^{(i)} \rho \sigma_{-}^{(j)} - \frac{1}{2} \{ \sigma_{-}^{(j)} \sigma_{+}^{(i)}, \rho \} \right), \label{eq1.4}
\end{aligned}
\end{equation}
where $\Gamma_{(ij)}^\pm$ are the dissipative rates corresponding to emission ($\Gamma_{(ij)}^+$) and absorption ($\Gamma_{(ij)}^-$) processes of energy quanta.

The dissipative rates can be determined by the bath structure and the dipole alignment between transitions \cite{Open1}. For simplicity, we assume the dipole moments $\mathbf{d}_{(1)}$ and $\mathbf{d}_{(2)}$ have unit magnitude, and denoting $\Theta_{ij}$ as the angle between them, the rates can be expressed as:
\begin{equation}
\Gamma_{(ij)}^\pm = \gamma_\pm \cos \Theta_{ij}, \label{eq:gamma}
\end{equation}
where $\gamma_\pm$ are the total downward (emission) and upward (absorption) rates, respectively, and can be calculated as the Fourier transforms of the bath correlation function. That is, for a homogeneous and isotropic thermal bath, the two-point function reduces to:
\begin{equation}
\langle \Phi_{\alpha}^{\dagger}(t) \Phi_{\beta}(t') \rangle = \delta_{\alpha\beta} \mathcal{G}(t - t'),
\label{eq:corl}
\end{equation}
where $\mathcal{G}(s)$ is the Wightman function of the bath field. The transition rates are then given by:
\begin{equation}
\gamma_{\pm}= \int_{-\infty}^{\infty} \mathrm{d}s\, e^{\pm i \omega s} \mathcal{G}(s).
\label{eq:spectral}
\end{equation}

One can confirm that {follows directly from the KMS condition, the transition rates $\gamma_\pm$ satisfy a detailed-balance relation. In fact, stationarity and Eq. \eqref{eq.3} imply $\mathcal{G}(s)=\mathcal{G}(-s-i \beta)$. Therefore, assuming the standard analyticity and decay properties that permit a contour shift in the complex plane $u=s+i \beta$, we have
\[
\gamma_{-} =\int_{-\infty}^{\infty} d s e^{-i \omega s} \mathcal{G}(-s-i \beta) 
 =e^{-\beta \omega} \int_{-\infty-i \beta}^{\infty-i \beta} d u e^{i \omega u} \mathcal{G}(u)\no
\]
Equivalently, this means that the transition rates satisfy}
\begin{equation}
\gamma_- = e^{-\beta \omega} \gamma_+, 
\label{eq:KMS}
\end{equation}
which manifests the KMS condition in the frequency domain.

For the \textsf{V}-type systems, the structure of the dissipator \eqref{eq1.4} of the master equation captures the coherent dynamics induced by degeneracy and indistinguishability of decay pathways. {In particular, while diagonal components $\Gamma_{(i i)}^{ \pm}$ produce the usual population relaxation and decoherence, the off-diagonal components $\Gamma_{(i j)}^{ \pm}$ ($i \neq j$) are cross-damping rates. Since the bath cannot distinguish the two aligned transitions, these terms couple populations to $\rho_{12}$ and can serve as a source of quantum coherence\footnote{{The mechanism is also clear in bright/dark basis $|B\rangle=(|1\rangle+|2\rangle) / \sqrt{2}$ and $|D\rangle=$ $(|1\rangle-|2\rangle) / \sqrt{2}$ \cite{Sufiani2026}. Two radiative branches to a common ground generate symmetry-adapted bright/dark superpositions, but the dark mode under decay realizes a decoherence-free subspace. Thermalization of the bright mode, together with preservation of the dark one, produces a population imbalance in the $\{|B\rangle,|D\rangle\}$ basis, which appears as a nonzero $\rho_{12}$ in the original transition basis. This explains how a dissipator can contain both decohering diagonal terms and coherence-generating cross terms.}
}.} This coherent cross-coupling, induced by interference, is a hallmark of V-type systems \cite{Tscherbul2014} and plays a central role in coherence-enhanced thermodynamic processes \cite{Vtype1,Min2025}. {Moreover, for perfectly aligned dipoles, $\cos \Theta_{i j}=1$ for all $i, j$, the interference between decay channels is maximized.}

In the next section, we show how the specific $\Gamma_{(i j)}^{ \pm}$ structure is responsible for generating and sustaining quantum coherence in the steady state, which is anticipated to be harnessed for thermodynamic tasks such as work extraction.


\subsection{Coherence generation with aligned dipoles}
\label{II.2}

To describe the density matrix of the three-level system, we use the Gell-Mann matrices (see Appendix \ref{appA} for notation) as a basis for arbitrary $3\times 3$ matrices. In a generalized Bloch form, the density matrix $\rho$ of a three-level system can be expanded as:
\begin{equation}
\rho = \frac{1}{3} \left( \mathbb{I}_{3 \times 3} + \sum_{i=1}^{8} q_i \mathbb{P}_i \right),
\label{eq:bloch}
\end{equation}
where the real vector $\mathbf{q} = (q_1, q_2, \ldots, q_8)$ is the generalized Bloch vector. In component form, the density matrix is
\be
\rho 
=\frac{1}{3}\left(\begin{array}{ccc}
1+q_7 & q_1 & q_3 \\
q_2 & 1+q_8 & q_5 \\
q_4 & q_6 & 1-q_7-q_8
\end{array}\right).
\label{eq:comp}
\ee

Substituting Eq.\eqref{eq:bloch} into the master equation Eq.\eqref{master}, we obtain a set of first-order differential equations for the Bloch vector components, which can further be divided into three decoupled sets. In particular, the time evolution of $\{ q_1, q_2, q_7, q_8 \}$ is governed by
\begin{equation}
\begin{aligned}
& \left\{
\begin{aligned}
\dot{q}_1 &= \gamma_{-} \mathfrak{p} \left[ 1 - (q_7 + q_8) \right]- \gamma_{+} \left[ q_1 + \mathfrak{p} \left( 1 + \frac{q_7 + q_8}{2} \right) \right] \\
\dot{q}_2 &= \gamma_{-} \mathfrak{p} \left[ 1 - (q_7 + q_8) \right]- \gamma_{+} \left[ q_2 + \mathfrak{p} \left( 1 + \frac{q_7 + q_8}{2} \right) \right] \\
\dot{q}_7 &=\gamma_{-} \left[ 1 - (q_7 + q_8) \right] - \gamma_{+} \left[ 1 + q_7 + \frac{\mathfrak{p}}{2} (q_1 + q_2) \right]  \\
\dot{q}_8 &= \gamma_{-} \left[ 1 - (q_7 + q_8) \right]- \gamma_{+} \left[ 1 + q_8 + \frac{\mathfrak{p}}{2} (q_1 + q_2) \right] 
\end{aligned}
\right.
\end{aligned}
\label{eqns-1}
\end{equation}
while the dynamics of $\{ q_3, q_5 \}$ are governed by
\begin{equation}
\begin{aligned}
& \left\{
\begin{aligned}
\dot{q}_3 &= -i \omega q_3  -\frac{\gamma_{+}}{2} \left( q_3 + \mathfrak{p} q_5 \right) - \gamma_{-} q_3 \\
\dot{q}_5 &= -i \omega q_5 -\frac{\gamma_{+}}{2} \left( q_5 + \mathfrak{p} q_3 \right) - \gamma_{-} q_5
\end{aligned}
\right.~~~~~~~~~~~~
\end{aligned}
\label{eqns-2}
\end{equation}
and for $\{ q_4, q_6 \}$, we have
\begin{equation}
\begin{aligned}
& \left\{
\begin{aligned}
\dot{q}_4 &= i \omega q_4 -\frac{\gamma_{+}}{2} \left(  q_4 + \mathfrak{p} q_6 \right) - \gamma_{-} q_4 \\
\dot{q}_6 &= i \omega q_6 -\frac{\gamma_{+}}{2} \left(  q_6 + \mathfrak{p} q_4 \right) - \gamma_{-} q_6
\end{aligned}
\right.~~~~~~~~~~~~~~
\end{aligned}
\label{eqns-3}
\end{equation}
Here, $\mathfrak{p}:= \cos \Theta_{12}$ characterizes the alignment between dipoles.

Since three sets of Bloch vector components are dynamically decoupled, it is instructive to consider the case with $q_3=q_4=q_5=q_6=0$. This restriction on the state space will not interfere with our search for coherence generation, as Eq.\eqref{eqns-2} and Eq.\eqref{eqns-3} only admit time-decay solutions. From Eq.\eqref{eq:comp}, we see then the master equation \eqref{eqns-1} about $\{ q_1, q_2, q_7, q_8 \}$ describes the dynamics of $\{ \rho_{21}, \rho_{12}, \rho_{22}, \rho_{11} \}$\footnote{The dynamics of $\rho_{00}$ is determined by $\text{Tr}\left(\rho(t)\right)=1$.}, encoding the relevant population and coherence, respectively.

Defining the (anti-)symmetric combination $\rho_{\pm} = (\rho_{21} \pm \rho_{12})/2$, the master equation \eqref{eqns-1} can be conveniently rewritten as a Schr\"odinger-like equation for the coherence vector $\bm{\Pi} =\left(\rho_{22}, \rho_{00}, \rho_{+}, \rho_{-}\right)^{\intercal}$ \cite{Alicki1987}: 
\begin{equation}
\dot{\bm{\Pi}} = \mathbf{M}\cdot\bm{\Pi} - \mathbf{b}, \label{eq.16}
\end{equation}
where
\begin{equation}
\mathbf{M} = \left(\begin{array}{cccc}
-\gamma_{+} & \gamma_{-} & -\mathfrak{p} \gamma_{+} & 0 \\
0 & -(\gamma_{+} + 2 \gamma_{-}) & 2 \mathfrak{p} \gamma_{+} & 0 \\
0 & \displaystyle\frac{\mathfrak{p}}{2} (\gamma_{+} + 2 \gamma_{-}) & -\gamma_{+} & 0 \\
0 & 0 & 0 & -\gamma_{+}
\end{array}\right),  \label{eq.17}
\end{equation}
and 
\be
\mathbf{b} = \left(0,-\gamma_{+},\frac{\mathfrak{p}}{2} \gamma_{+},0\right)^{\intercal}.
\ee

For $|\mathfrak{p}| \neq 1$, the coefficient matrix $\mathbf{M}$ in Eq.\eqref{eq.16} is invertible. The real parts of its eigenvalues are negative, so that the system eventually reaches the Gibbs state \cite{Lendi1987} 
\be
\rho(\infty)=\text{diag}\left(\frac{1}{2+e^{\beta\omega}},\frac{1}{2+e^{\beta\omega}},\frac{e^{\beta\omega}}{2+e^{\beta\omega}}\right)
\ee

%

For $|\mathfrak{p}| = 1$, aligned dipole moments satisfy $\mathbf{d}_{(1)}=\mathbf{d}_{(2)}$ and the matrix $\mathbf{M}$ becomes irreversible. Taking $\mathfrak{p} = 1$, for example, Eq.\eqref{eq.16} can be  analytically resolved. Denoting the initial conditions as $\rho_{22}(0) = a, ~ \rho_{00}(0) = b, ~ \rho_{+}(0) = c, ~ \rho_{-}(0) = i d$, with $a, b, c, d \in \mathbb{R}$ and defining $x:= e^{-\beta \omega}$, the time evolution of the relevant density matrix components reads from Eq.\eqref{eqns-1} as:
\begin{equation}
 \left\{\begin{aligned}
\rho_{22}(t) &= \frac{1}{4(1+x)} \left\{ 2(1 + x)(2a + b - 1) e^{-\gamma_+t}\right. \\
&~~~~~~~~~~~~~+ \left[1 + 2c - (1 + 2x)b \right] e^{-2(1+x)\gamma_+t} \\
&~~~~~~~~~~~~~+\left.(1 + 2x - b - 2c)\right\}, \\
\rho_{00}(t) &= \frac{1}{2(1+x)} \{ \left[ -1 - 2c + (1 + 2x)b \right] e^{-2(1+x)\gamma_+t}\\
&~~~~~~~~~~~~~\left.+(1 + b + 2c) \right\}, \\
\rho_{12}(t) &= \frac{1}{4(1+x)} \{\left[ 1 + 2c - (1 + 2x)b \right] e^{-2(1+x)\gamma_+t}\\
&~~~~~~~~~~~~~\left.+(1 + 2x)(b + 2c) -1 \right\} - i d e^{-\gamma_+t},
\end{aligned}\right.
\label{eqq}
\end{equation}
while $\rho_{11}=1-\rho_{22}-\rho_{00}$, attributed to trace preservation of density matrix $\operatorname{Tr}(\rho)=1$.

We immediately observe from Eq.\eqref{eqq} that besides the time decay terms, time-independent terms exist for non-diagonal components of the density matrix. This indicates that in the long-time limit, the steady state of the three-level system may retain memory of its initial coherence. In particular, we take the limit $t \rightarrow \infty$ on Eq.\eqref{eqq}, and obtain the asymptotic steady-state of three-level system as:
\begin{equation}
 \left\{\begin{aligned}
\rho_{22}(\infty) &= \frac{1 + 2 e^{-\beta \omega} - b - 2c}{4(1 + e^{-\beta \omega})}, \\
\rho_{00}(\infty) &= \frac{1 + b + 2c}{2(1 + e^{-\beta \omega})}, \\
\rho_{12}(\infty) &=\frac{-1 + (1 + 2 e^{-\beta \omega})(b + 2c)}{4(1 + e^{-\beta \omega})}.
\end{aligned}\right.
 \label{eq.20}
\end{equation}
The nonvanishing off-diagonal element $\rho_{12}$ suggests the possibility of non-thermal steady states that contain residual coherence. To see this, we use the so-called $\ell_1$-norm
\be
C_{\ell_1}(\rho)=\sum_{i, j; i \neq j}\left|\rho_{i, j}\right|,
\ee
a monotone measure of quantum coherence in the framework of resource theory \cite{Baumgratz_2014}, to quantify the coherence of the \textsf{V} atom.

{We are particularly interested in the asymptotic state because it is physically useful, as it is stationary and provides a reproducible input resource for the work-extraction protocols discussed later.} For the asymptotic steady-state $\rho_\infty$ given by Eq.\eqref{eq.20}, it inherits quantum coherence
\be
C_{\ell_1}(\rho_\infty)= \frac{ \left| (1 + 2 e^{-\beta \omega})(b + 2c)-1\right|}{2(1 + e^{-\beta \omega})}
\ee
Notably, even if the three-level system is prepared on the ground level (i.e., $b=1$, $a=c=d=0$), the final $\ell_1$-norm of quantum coherence is nonzero:
\begin{equation}
C_{l_1}(\rho_{\infty})=\sum_{\substack{i, j \\ i \neq j}}\left|\rho_{i j}\right| = \frac{1}{e^{\beta \omega} + 1} \neq 0,
\end{equation}
determined by the bath temperature and transition energy of the excited states.

In summary, we conclude that when there is dipole alignment ($\mathfrak{p}=1$), as well as for dipole anti-alignment ($\mathfrak{p}=-1$), the perfectly degenerate three-level system maintains long-lived quantum coherence induced by the atom-bath interaction. This phenomenon occurs because the two transition pathways (from excited states to the ground state in Fig.\ref{Degenerate 3-level}) become indistinguishable, resulting in quantum coherence generation between $\ket{1}$ and $\ket{2}$, even without external driving. Such long-lived quantum coherence makes the \textsf{V}-type three-level system a valuable device for energy extraction from quantum coherence.

\subsection{Nearly degenerate three-level system}

\begin{figure}[hbtp]
\centering  
\includegraphics[width=0.3\textwidth]{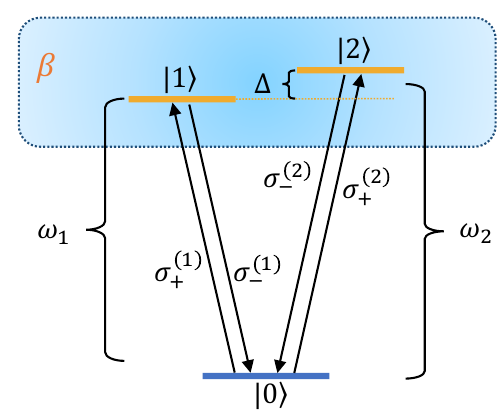}
\caption{Scheme of a nearly degenerate three-level atom. {The excited states $\ket{1}$ and $\ket{2}$ have a tiny energy splitting $\delta = \omega_2 - \omega_1$. Both dipole-allowed transitions couple to the same thermal reservoir at inverse temperature $\beta$.}} 
\label{Nearly degenerate 3-level}
\end{figure}

We now consider a more relaxed scenario, namely a nearly degenerate three-level \textsf{V}-type open quantum system, where the excited states $\ket{1}$ and $\ket{2}$ have slightly different energies (see Fig.\ref{Nearly degenerate 3-level}). The system Hamiltonian becomes
\begin{equation}
H_S = \omega_1 \ket{1}\bra{1} + \omega_2 \ket{2}\bra{2}, \label{eq2.13}
\end{equation}
where the energy splitting $\delta = \omega_2 - \omega_1$ is assumed to be tiny, with $\delta \ll \omega_i$ ($i=1,2$).

Compared to the degenerate case, the presence of energy splitting $\delta$ suggests that we should analyze the system dynamics within a metastable timescale over $\tau_S \ll t \ll 1/\delta$, where $\tau_S \sim 1/\omega_1$ is the characteristic relaxation timescale of the atom. On the other hand, in deriving the master equation, the standard secular approximation must be modified \cite{Tscherbul2014}, since slowly oscillating terms with $e^{\pm i (\omega_1 - \omega_2)t}$ should be retained. Following the standard procedure of microscopic derivation of the master equation \cite{Open1}, we can derive the complete dissipator $\mathcal{L}_2[\tilde{\rho}]$ for the general non-degenerate three-level open system (see Appendix~\ref{Appendix A} for details). In the degenerate limit, one has $\mathcal{L}_2[\tilde{\rho}]\xlrar{\delta = 0}\mathcal{L}_1[\rho]$, recovering the previously degenerate dissipator \eqref{eq1.4}.


To resolve the open dynamics, we simplify the problem by focusing on Bloch vector components $\{ \tilde{q}_1, \tilde{q}_2, \tilde{q}_7, \tilde{q}_8 \}$. The remaining components of the density matrix are dynamically decoupled from the coherence vector and may consistently be set to zero. In analogy with the degenerate scenario Eq.\eqref{eq.16}, we define the dressed combinations:
\begin{equation}
\tilde{\rho}_{\pm} = \frac{1}{2} \left( e^{-i \delta t} \tilde{\rho}_{21} \pm e^{i \delta t} \tilde{\rho}_{12} \right),
\end{equation}
and deduce the Schr\"odinger-like equation for the coherence vector $\boldsymbol{\widetilde{\Pi}}=\left( \tilde{\rho}_{22},  \tilde{\rho}_{00}, \tilde{\rho}_{+}, \tilde{\rho}_{-}\right)^{\top}$ as
\begin{equation}
\dot{\widetilde{\bm{\Pi}}} = \widetilde{\mathbf{M}}\cdot\widetilde{\bm{\Pi}} - \mathbf{b}, \label{eq.nd}
\end{equation}
where
\begin{widetext}
\begin{equation}
\widetilde{\mathbf{M}}= 
\left(\begin{array}{cccc}
-\gamma_{+,(2)} & \gamma_{-,(2)} & -\mathfrak{p} \gamma_{+,(1)} & 0 \\
\gamma_{+,(2)} - \gamma_{+,(1)} & -(\gamma_{+,(1)} + \gamma_{-,(1)} + \gamma_{-,(2)}) & \mathfrak{p}(\gamma_{+,(1)} + \gamma_{+,(2)}) & 0 \\
\displaystyle \frac{\mathfrak{p}}{2} (\gamma_{+,(1)} - \gamma_{+,(2)}) & \displaystyle \frac{\mathfrak{p}}{2}(\gamma_{+,(1)} + \gamma_{-,(1)} + \gamma_{-,(2)}) & \displaystyle -\frac{1}{2}(\gamma_{+,(1)} + \gamma_{+,(2)}) & -i\delta \\
0 & 0 & -i\delta & -\frac{1}{2}(\gamma_{+,(1)} + \gamma_{+,(2)})
\end{array}\right)
\label{eqMnd}
\end{equation}
\end{widetext}
with 
\begin{equation}
\gamma_{\pm,(i)}= \int_{-\infty}^{\infty} \mathrm{d}s\, e^{\pm i \omega_i s} \mathcal{G}(s).
\label{eq:spectral1}
\end{equation}
for $i=1,2$ and same $\mathbf{b}$ in Eq.(\ref{eq.17}). 

Eq.\eqref{eq.nd} is a set of first-order equations that can, in principle, be resolved analytically, although the explicit solution is prohibitively lengthy. Since the energy splitting $\delta$ is tiny, we can alternatively present a perturbative solution to Eq.\eqref{eq.nd} in Appendix.\ref{appendixc}. {Unlike the degenerate \textsf{V} atom, the presence of $\delta$ makes the noise-induced coherence survive for an extremely long time \cite{Tscherbul2014}, but not a genuine coherent asymptotic steady state.}

\section{Work Extraction from Long-Lived Quantum Coherence} 
\label{sec3}

We now study how the long-lived quantum coherence of a three-level \textsf{V} atom is used in explicit work-extraction protocols \cite{Korzekwa2016}. To proceed, we first introduce the concept of {maximal extractable work} in Section \ref{3.1}, quantified by the so-called {free energy difference} \cite{Brandao2013}. Then, a concrete work-extraction protocol (\textbf{Protocol-1}) is proposed in Section \ref{protocol1}. It starts from the coherent asymptotic state of the exactly degenerate dynamics and converts its coherence into practical work through a repeated cycle. Finally, in Section \ref{protocol2}, we refine the scheme to a one-cycle quasistatic construction (\textbf{Protocol-2}) that reaches the bath-assisted FED bound.

 \subsection{Free energy difference}
 \label{3.1}

The \emph{free energy difference} (FED) is a benchmark of the maximum average work extractable from a quantum system in contact with a thermal reservoir \cite{Goold2016}. Let $\rho$ be the quantum state of a system with Hamiltonian $H_S$, which interacts with a heat bath at inverse temperature $\beta = 1/(kT)$. The \emph{maximum extractable work} (MEW)\footnote{{In addition to the FED, various benchmarks for extractable work exist. A widely used one is ergotropy $\mathcal{E}(\rho)=\operatorname{Tr}\left(H_S \rho\right)-\min _U \operatorname{Tr}\left(H_S U \rho U^{\dagger}\right)$ for quantum batteries \cite{Allahverdyan2004}. The appropriate choice depends on the specific operations under discussion. For example, while ergotropy generally characterizes isolated cyclic unitaries \cite{Cakmak2020}, the FED gives the maximum average bath-assisted work extraction, which we used in protocols. Nevertheless, one in general has $\Delta F_\beta(\rho) \geq \mathcal{E}(\rho) $, which is saturated when a passive state equals the Gibbs state at the bath temperature.}} via thermal operations within a coherent reference \cite{Korzekwa2016} is bounded by the FED between the system's state $\rho$ and a Gibbs state $\gamma_S$ \cite{Brandao2013}:
\begin{equation}
\langle W \rangle_\rho^{\max} = \Delta F_\beta(\rho) := F_\beta(\rho) - F_\beta(\gamma_S)
\label{FED}
\end{equation}
Here the non-equilibrium free energy is defined as $F(\sigma) = \operatorname{Tr}(\sigma H_S) - T S(\sigma)$, and $S(\sigma) = -\operatorname{Tr}(\sigma \log \sigma)$ is the von Neumann entropy. The Gibbs state $\gamma_S$ is chosen as
\begin{equation}
\gamma_S = \frac{e^{-\beta H_S}}{\mathcal{Z}_S},~~~\mathcal{Z}_S:=\operatorname{Tr}(e^{-\beta H_S})
\label{Gibbs}
\end{equation}
which gives $F(\gamma_S) = -T \log \mathcal{Z}_S$ and has obviously no work can be extracted further.

Applying the FED \eqref{FED} to the scenario of three-level system with energy eigenstates $\{\ket{0}, \ket{1}, \ket{2}\}$. For example, we consider the state subspace spanned by $\{ \rho_{21}, \rho_{12}, \rho_{22}, \rho_{11} \}$ as before. Three eigenvalues of such density matrix are $\lambda_0=1-\rho_{11}-\rho_{22}$, and 
\be
\lambda_\pm= \frac{1}{2} \left[ \rho_{11} + \rho_{22} \pm \sqrt{(\rho_{11} - \rho_{22})^2 + 4 |\rho_{12}|^2} \right].
\label{eigenvalues}
\ee

With the degenerate system Hamiltonian \eqref{eq.1}, the related FED can be directly calculated as:
\begin{equation}
 \Delta F_\beta(\rho)  = \omega( \rho_{11} +  \rho_{22})
+ kT \left[ \sum_{i=0,\pm} \lambda_i \log \lambda_i + \log{\mathcal{Z}_S} \right],
\label{work}
\end{equation}
where $\mathcal{Z}_S = 1 + 2e^{-\beta \omega}$. 

The interesting point is that quantum coherence (via $\rho_{12}$) reduces the von Neumann entropy term, thereby allowing for more MEW (indicated by the larger FED \eqref{work}) to be potentially utilized. This clearly provides a quantum advantage over classical thermodynamics, which only considers populations.

{It is insightful to recast the FED via quantum coherence. By introducing the total excited population $\mathcal{P}_{\text{exc}}:=\rho_{11}+\rho_{22}$, the population asymmetry $\mathcal{P}_{\text{asym}}=:|\rho_{11}-\rho_{22}|$, and coherence $\ell_1$-norm $C_{\ell_1}(\rho)=2\left|\rho_{12}\right|$, the eigenvalues \eqref{eigenvalues} can be reformulated as:
\bea
\lambda_0&=&1-\mathcal{P}_{\text{exc}} \no\\
\lambda_{ \pm}&=&\frac{1}{2}\left[\mathcal{P}_{\text{exc}} \pm \sqrt{\left(\mathcal{P}_{\text{asym}}\right)^2+C_{\ell_1}^2}\right]
\label{evalues}
\eea
Substituting them into Eq. \eqref{work}, we obtain
\bea
\Delta F_\beta(\rho)&=&\omega \mathcal{P}_{\text{exc}}+\beta^{-1}\log{\mathcal{Z}_S}\no\\
&&-\beta^{-1}\left[ H_{\text{bin}}\left(\mathcal{P}_{\text{exc}}\right)+\mathcal{P}_{\text{exc}}H_{\text{bin}}\left(\frac{\lambda_{+}}{\mathcal{P}_{\text{exc}}}\right)\right]~~~~~
\label{FED-C}
\eea
where $ H_{\text{bin}}(x):=-x \log x-(1-x) \log (1-x)$ is binary entropy function. The entropic terms in Eq.\eqref{FED-C} have an intuitive interpretation. Specifically, $H_{\text{bin}}\left(\mathcal{P}_{\text{exc}}\right)$ characterizes the ground-excited population distribution, whereas $\mathcal{P}_{\text{exc}}H_{\text{bin}}\left({\lambda_{+}}/{\mathcal{P}_{\text{exc}}}\right)$ captures the excited-subspace distribution, encoding the influence of noise-induced quantum coherence.

In Fig.\ref{generalFEDmap-1}, we plot FED as a function of quantum coherence $C_{\ell_1}$ for various population configurations, using the level spacing $\omega$ and inverse temperature $\beta$ as control parameters. In particular, Figs.\ref{generalFEDmap-1}(a), (b) depict symmetric populations $\left(\mathcal{P}_{\text {asym }}=0\right)$, while Figs.\ref{generalFEDmap-1}(c), (d) show asymmetric cases $\left(\mathcal{P}_{\text {asym }}=0.2\right)$. The estimation is performed over the physical bound $0 \leq C_{\ell_1} \leq 2 \sqrt{\rho_{11} \rho_{22}}$, ensuring the positivity of the density matrix.

\begin{figure}[hbtp]
\centering
\includegraphics[width=0.5\textwidth]{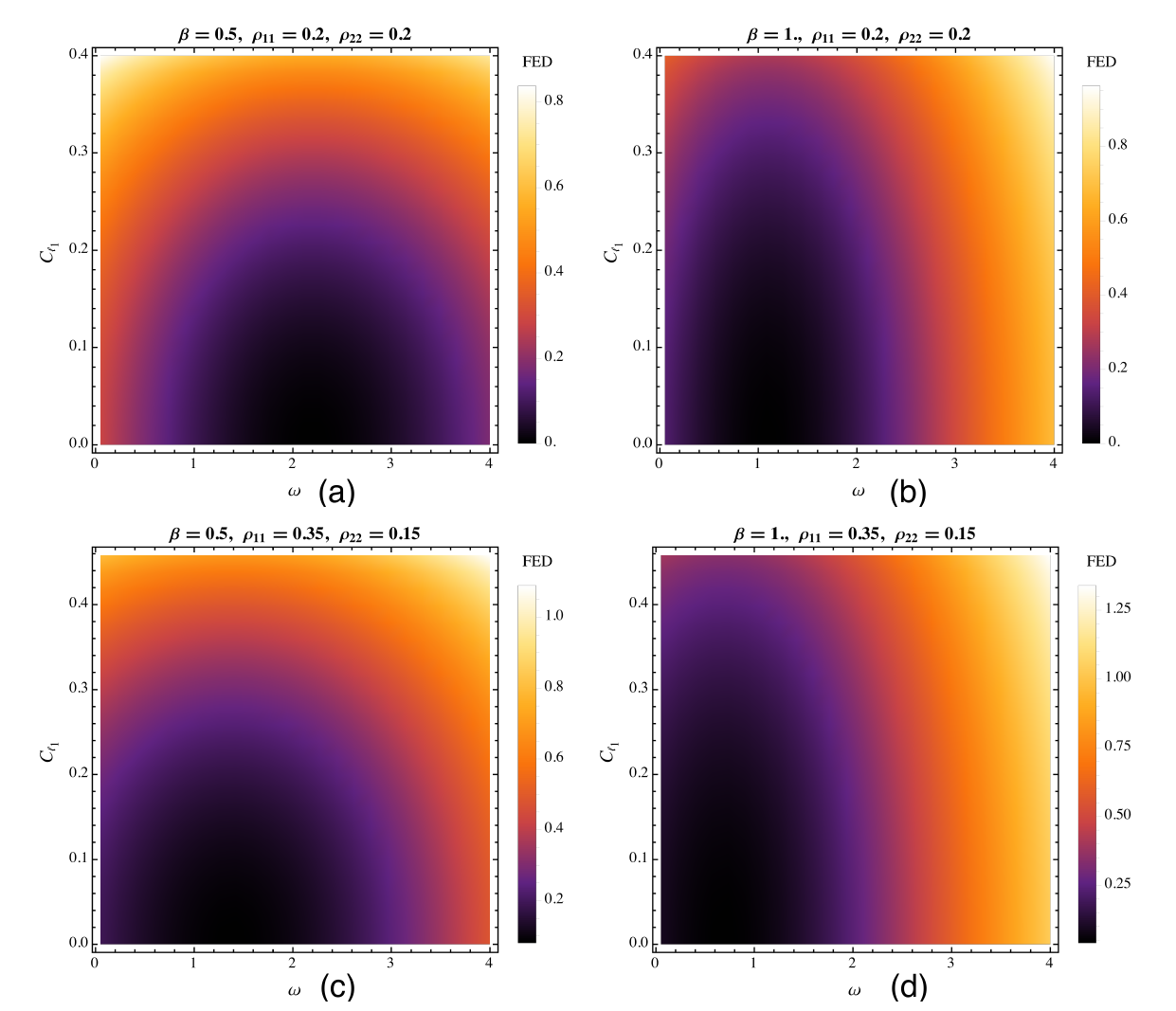}
\caption{{The FED as a function of $\ell_1$-norm $C_{\ell_1}(\rho)$. The four panels set the state with symmetric population in (a) $(\rho_{11},\rho_{22})=(0.20,0.20)$, $\beta=0.5$, and (b) $(\rho_{11},\rho_{22})=(0.20,0.20)$, $\beta=1$, while asymmetric population is setted in (c) $(\rho_{11},\rho_{22})=(0.35,0.15)$, $\beta=0.5$, and (d) $(\rho_{11},\rho_{22})=(0.35,0.15)$, $\beta=1$, respectively. 
}}
\label{generalFEDmap-1}
\end{figure}

We observe that at fixed $\omega$, the FED increases monotonically with coherence $C_{\ell_1}$, consistent with Eq.\eqref{FED-C}. That is, quantum coherence amplifies the separation between excited-block eigenvalues $\lambda_{ \pm}=\frac{1}{2}\left[\mathcal{P}_{\text {exc }} \pm \sqrt{\mathcal{P}_{\text {asym }}^2+C_{\ell_1}^2}\right]$, thereby suppressing the entropic terms and increasing the FED. In short, enhanced coherence provides more free energy for extraction.

Second, a comparison between Figs.\ref{generalFEDmap-1}(a), (c) with $\beta=0.5$ and Figs.\ref{generalFEDmap-1}(b), (d) with $\beta=1$ reveals that lower temperatures reduce the overall magnitude of the FED and shift its minimum to smaller $\omega$. This can be attributed to the thermal contribution scaling as $\beta^{-1}$ to the FED, and at lower temperatures, the suppressed thermal occupation of excited states causes the system to deviate less from the Gibbs state.

Finally, contrasting Figs.\ref{generalFEDmap-1}(a), (b) with Figs.\ref{generalFEDmap-1}(c), (d) demonstrates that a nonvanishing population asymmetry (i.e., $\mathcal{P}_{\text {asym }}=0.2$) significantly reshapes the FED distribution. In Figs.\ref{generalFEDmap-1}(a), (b), the FED minimum along the $\omega$ axis occurs when the total excited population matches the Gibbs distribution (e.g., $\mathcal{P}_{\mathrm{exc}}^{\mathrm{eq}}=\frac{2 e^{-\beta \omega}}{1+2 e^{-\beta \omega}}$ and  $\mathcal{P}_{\text {asym }}^{\mathrm{eq}}=0$ for $\gamma_S$). Once, in addition, $C_{\ell_1}=0$, the system is in thermal equilibrium, and the FED vanishes. In Figs.\ref{generalFEDmap-1}(c), (d), however, the persistent population asymmetry prevents the FED from vanishing entirely. This highlights that the FED is governed jointly by the total excited population $\mathcal{P}_{\text {exc }}$, the population asymmetry $\mathcal{P}_{\text {asym}}$, and quantum coherence.

}

\subsection{The work extraction protocol}
\label{protocol1}

{
Without loss of generality, we assume the exactly degenerate \textsf{V} atom starts in its ground state $\ket{0}\bra{0}$ (parameterized by $a=c=d=0$ and $b=1$). Substituting this into the asymptotic steady state \eqref{eq.20} yields a specific form}:
\begin{equation}
\rho_0 =\frac{1}{1+e^{\beta\omega}}
\left(\begin{array}{ccc}
1/2 & 1/2 & 0 \\
1/2 & 1/2 & 0 \\
0 & 0 & e^{\beta\omega}
\end{array}\right),
\label{eq.34}
\end{equation}
with nonvanishing noise-induced quantum coherence. {We treat the stationary coherence of the asymptotic state \eqref{eq.34} as a resource and aim to design an optimal protocol for converting it into useful work. }


{Noting from Eqs.\eqref{evalues} that quantum coherence $C_{\ell_1}$ and population asymmetry $\mathcal{P}_{\text {asym }}$ play equivalent roles in determining the FED \eqref{FED-C}. Given that population manipulation is more amenable to experiment, our protocol is designed to first completely transform the coherence into population asymmetry, after which useful work is extracted through controlled level modulation. }

To do so, we introduce a unitary operator:
\begin{equation}
\mathcal{U} = \frac{1}{\sqrt{2}} 
\left(\begin{array}{ccc}
1 & -1 & 0 \\
1 & 1 & 0 \\
0 & 0 & \sqrt{2}
\end{array}\right),
\label{epreserve}
\end{equation}
Since $[\mathcal{U},H_S]=0$, the operation can be regarded as an excited-subspace rotation that preserves the system energy. Nevertheless, under this energy-preserving operation, we can show later that quantum coherence between degenerate energy levels could be converted to an asymmetry between their populations.

\begin{figure}[htbp]
\centering  
\includegraphics[width=0.5\textwidth]{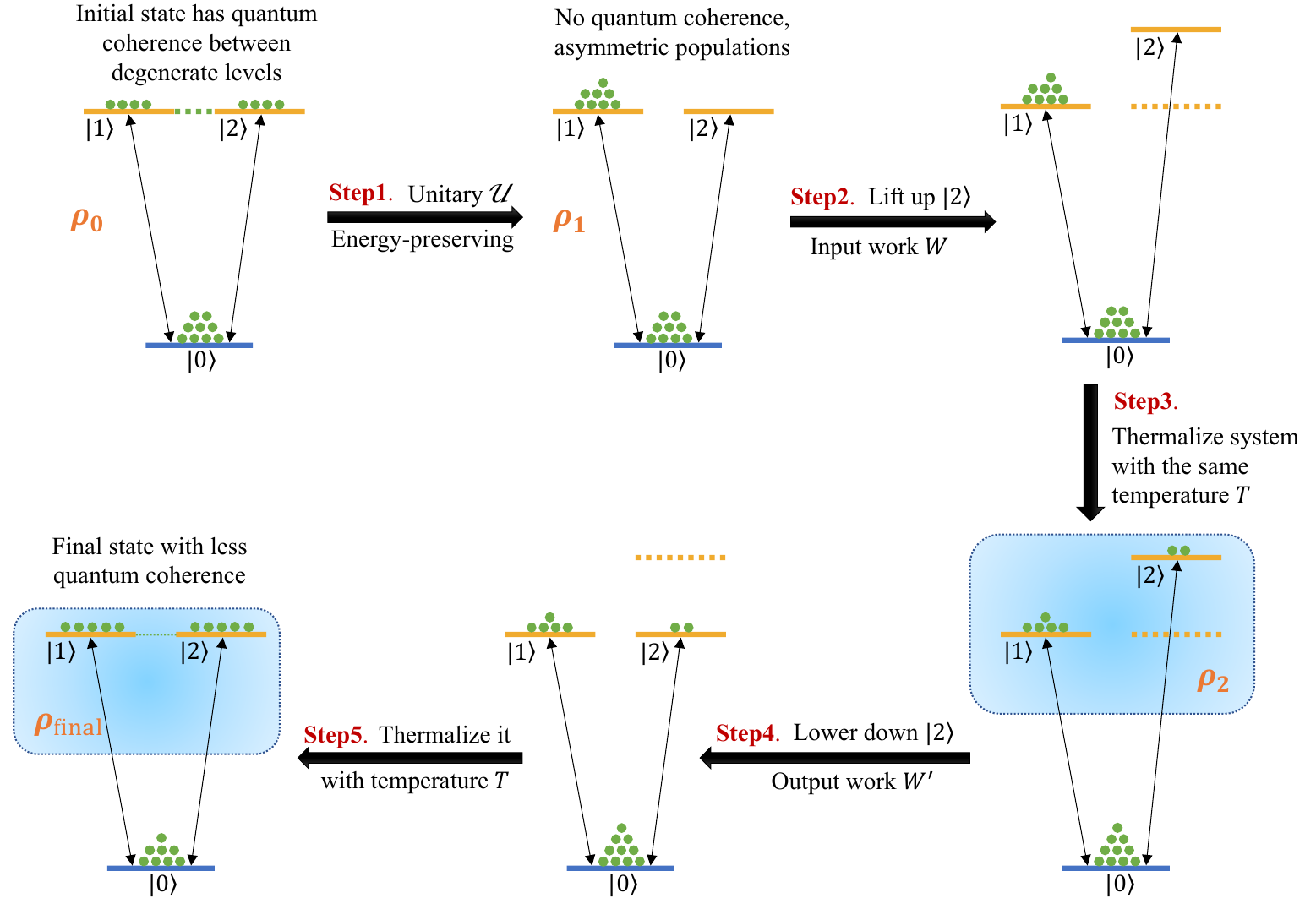}
\caption{Schematic of the work extraction Protocol-1 from a degenerate three-level quantum system interacting with a thermal bath. The protocol starts with a coherent steady state induced by environmental interaction, then applies an energy-preserving unitary transformation that converts coherence into population asymmetry. Next, a series of adiabatic level shifts and thermalizations is used to extract work from the asymmetrically populated states. The process can be repeated multiple times to use up residual coherence and maximize work output.} \label{Protocol}
\end{figure}

With the above setting, we now outline the steps of the work-extraction \textbf{Protocol-}$\mathbf{1^{(1)}}$\footnote{Since Protocol-1 can be repeated, we use the superscript ${(n)}$ to denote the $n$-th round of protocol execution.}, as illustrated schematically in Fig.\ref{Protocol}:\\

\noindent\textbf{Step 1}: Apply the unitary operation $\mathcal{U}$ to the asymptotic steady state $\rho_0$, resulting in the system arriving at:
 \be
    \rho_1^{(1)} = \mathcal{U} \rho_0 \mathcal{U}^\dagger=\frac{e^{-\beta\omega}}{1+e^{-\beta\omega}}\left(\begin{array}{ccc}
0 & 0 & 0 \\
0 & 1 & 0 \\
0 & 0 & e^{\beta\omega}
\end{array}\right),
    \ee
with population at $\ket{1}$ equal to $e^{-\beta \omega}/(1 + e^{-\beta \omega})$, and $\ket{2}$ is unpopulated. Thus, $\rho_1^{(1)}$ is diagonal in the new frame, and the initial coherence has been converted entirely into excited-state population asymmetry.
    
\noindent\textbf{Step 2}: Adiabatically raise the energy of $\ket{2}$ to $\Omega^{(1)}:={\omega} + {\tilde{\omega}}^{(1)}$, so that the system Hamiltonian becomes $H_S^{(1)}=\omega \ket{1}\bra{1} + \Omega^{(1)} \ket{2}\bra{2}$. Since $\ket{2}$ is unoccupied, no work is required during this elevation step, i.e., $W^{(1)} = 0$.
    
\noindent\textbf{Step 3}: Bring the non-degenerate system $\rho_1^{(1)}$ into contact again with an \emph{ancillary} heat bath at temperature $1/\beta$ and thermalize it. The open dynamics of this three-level system are described in the Appendix.\ref{Appendix A}. After a sufficiently long time, the non-degenerate three-level system behaves as two independent two-level open quantum systems that approach Gibbs states via thermalization governed by the dissipator \eqref{B2}. Eventually, the system arrives at the new equilibrium:
    \be
   \rho_2^{(1)}=\frac{1}{\mathcal{Z}^{(1)}} 
   \left(\begin{array}{ccc}
e^{-\beta\Omega^{(1)}} & 0 & 0 \\
0 & e^{-\beta\omega} & 0 \\
0 & 0 & 1
\end{array}\right),
\label{eqthermal}
    \ee
with $\mathcal{Z}^{(1)}:=1+e^{-\beta \omega}+e^{-\beta \Omega^{(1)}}$, indicating the population of $\ket{2}$ changes as
    \begin{equation}
  0~\lrar~  \frac{e^{-\beta\Omega^{(1)}}}{\mathcal{Z}^{(1)}}.
  \label{2population}
    \end{equation}
    
\noindent\textbf{Step 4}: Remove the ancillary bath, then let the system isolate again. We gradually lower the raised level $\ket{2}$ back to its original energy $\omega$ \cite{Frenzel2014}. Since the level $\ket{2}$ is now populated as \eqref{2population}, this step extracts a positive amount of work, which is:
    \begin{equation}
    \Delta W^{(1)}= \operatorname{Tr}\left(\rho_2^{(1)} H_S^{(1)}\right)- \operatorname{Tr}\left(\rho_2^{(1)} H_S\right)
    =\tilde{\omega}^{(1)} \frac{e^{-\beta \Omega^{(1)}}}{\mathcal{Z}^{(1)}}
       \end{equation}
    
\noindent\textbf{Step 5}: Finally, incorporate the ancillary bath interaction again and thermalize $\rho_2^{(1)}$ under the same dipole-type interaction at the same temperature $1/\beta$. The resulting steady density matrix is denoted as $\rho_{\text{final}}^{(1)}$, given by Eq.\eqref{eq.D2} in Appendix.\ref{AppendixD}.\\

We conclude that the net work extracted from conducting Protocol-1 is: $W_{\text{protocol-1}}^{(1)}= \Delta W^{(1)} - W^{(1)} = \Delta W^{(1)}$, which reaches its maximum when the optimal $\tilde{\omega}^{(1)}$ is obtained from $d W^{(1)}_{\text{protocol-1}}/d\tilde{\omega}^{(1)}=0$, as:
\begin{equation}
\tilde{\omega}^{(1)}_{\text{optimal}}= \frac{1}{\beta} \left[ 1 + \mathcal{W}_p\left( \frac{1}{\left(1 + e^{\beta \omega}\right) e} \right) \right].
\label{optimal}
\end{equation}
Here $\mathcal{W}_p(z)$ is the principal branch of the Lambert $\mathcal{W}$-function \cite{Olver2010}, satisfying $\mathcal{W} e^\mathcal{W} = z$. 

We note that the quantum coherence arising from degeneracy serves as a physical resource that is gradually consumed during the work-extraction process. To see this, we can compare the quantum coherence initially possessed in $\rho_0$ \eqref{eq.34}, i.e., $C_{l_1}(\rho_0)= (1+e^{\beta \omega})^{-1}$, to the quantum coherence of the final state \eqref{eq.D2}:
\begin{equation}
C_{l_1}(\rho^{(1)}_{\text{final}}) = C_{l_1}(\rho_0)\frac{ \left|1 - e^{-\beta \tilde{\omega}^{(1)}}\right|}{2\mathcal{Z}^{(1)}} < C_{l_1}(\rho_0).
\end{equation}
The decrement of quantum coherence is a trade-off for the work extracted by the protocol. 

However, since $\rho^{(1)}_{\text{final}}$ is still coherent after completing Protocol-1, we can continue extracting work by repeating the protocol until no more quantum coherence can be utilized, i.e., the final state after the protocol is a Gibbs state $\gamma_S$. 

For example, starting from the first round's final state, $\rho^{(1)}_{\text{final}}$, we initiate the second round of Protocol-1, called \textbf{Protocol-}$\mathbf{1^{(2)}}$, as follows:\\

\noindent\textbf{Step} $\mathbf{1^{(2)}}$: Perform an energy-preserving unitary transformation on state \eqref{eq.D2}, i.e., $ \rho_1^{(2)} = \mathcal{U} \rho^{(1)}_{\text{final}}\mathcal{U}^{\dagger}$, we obtain
     \be
   \rho_1^{(2)} = \left(\begin{array}{ccc}
\frac{\mathcal{Z}^{(1)}+e^{-\beta\tilde{\omega}^{(1)}}}{2(1+e^{\beta\omega})\mathcal{Z}^{(1)}}
 & 0 & 0 \\
0 & \frac{1+\mathcal{Z}^{(1)}}{2(1+e^{\beta\omega})\mathcal{Z}^{(1)}} & 0 \\
0 & 0 & \frac{(1+\mathcal{Z}^{(1)})e^{\beta\omega}}{2(1+e^{\beta\omega})\mathcal{Z}^{(1)}}
\end{array}\right),
    \ee
which is a diagonal state with asymmetric excited-state populations. In particular, the population of $\ket{2}$ becomes $\left(e^{-\beta\omega}+e^{-\beta\Omega^{(1)}}\right)/2\mathcal{Z}^{(1)}$.

\noindent\textbf{Step} $\mathbf{2^{(2)}}$: Lift the energy level of $\ket{2}$ adiabatically to $\omega + \tilde{\omega}^{(2)}$. The system Hamiltonian becomes $H_S^{(2)}=\omega \ket{1}\bra{1} + \Omega^{(2)} \ket{2}\bra{2}$, accompanied by work consumption: 
    \be
    W^{(2)} =\tilde{\omega}^{(2)} \frac{e^{-\beta{\omega}}+e^{-\beta\Omega^{(1)}}}{2\mathcal{Z}^{(1)}}
    \ee

\noindent\textbf{Step} $\mathbf{3^{(2)}}$: Bring the system into contact with ancillary heat bath at temperature $1/\beta$ to rethermalize the system. The population of $\ket{2}$ becomes
    \be
\frac{e^{-\beta{\omega}}+e^{-\beta\Omega^{(1)}}}{2\mathcal{Z}^{(1)}}~\lrar~ 
    \frac{e^{-\beta \Omega^{(2)}}}{\mathcal{Z}^{(2)}}.
    \ee
    where shifted energy level is $\Omega^{(2)}:=\omega+\omega^{(2)}$, and $\mathcal{Z}^{(2)}:=1+e^{-\beta\omega}+e^{-\beta\Omega^{(2)}}$.

\noindent\textbf{Step} $\mathbf{4^{(2)}}$: Lower the energy level of $\ket{2}$ gradually back toward $\omega$. The work extracted by this step is:
        \begin{equation}
    \Delta W^{(2)}= \operatorname{Tr}\left(\rho_2^{(2)} H_S^{(2)}\right)- \operatorname{Tr}\left(\rho_2^{(2)} H_S\right)
      =\tilde{\omega}^{(2)} \frac{e^{-\beta \Omega^{(2)}}}{\mathcal{Z}^{(2)}}
       \end{equation}

\noindent\textbf{Step} $\mathbf{5^{(2)}}$: Thermalize the system again under the same dipole interaction with the bath temperature $1/\beta$. The final density matrix, denoted as $\rho_{\text{final}}^{(2)}$, has the same form as $\rho_{\text{final}}^{(1)}$ with all $\tilde{\omega}^{(1)}$ replaced by $\tilde{\omega}^{(2)}$.\\

The net work extracted in Protocol-$1^{(2)}$ is
\be
W_{\text{protocol-1}}^{(2)}= 
\tilde{\omega}^{(2)}\left( \frac{e^{-\beta \Omega^{(2)}}}{\mathcal{Z}^{(2)}}-
\frac{e^{-\beta{\omega}}+e^{-\beta\Omega^{(1)}}}{2\mathcal{Z}^{(1)}}\right).
\ee
To ensure positive work extraction in the protocol, i.e., $W_{\text{protocol-1}}^{(2)}\geqslant0$, the energy level shift should satisfy
\begin{equation}
0 < \tilde{\omega}^{(2)} < \frac{1}{\beta} \log \left[ \frac{2 + e^{-\beta \omega} + e^{-\beta \Omega^{(1)}}}{(1 + e^{-\beta \omega})(1 + e^{-\beta \tilde{\omega}^{(1)}})} \right] < \tilde{\omega}^{(1)}.
\label{constr}
\end{equation}

Following the same strategy, we can repeatedly perform Protocol-$1$ and convert the remaining quantum coherence from the last cycle into additional work. The net work extraction in arbitrary round of Protocol-$1^{(i)}$ ($i=1,2,\cdots,n$) can be obtained: 
\be
\left\{
\begin{aligned}
W_{\text{protocol-1}}^{(i= 1)}&=\tilde{\omega}^{(1)}\left[ \frac{e^{-\beta \Omega^{(1)}}}{\mathcal{Z}^{(1)}}\right]\\
W_{\text{protocol-1}}^{(i\neq 1)}&=\tilde{\omega}^{(i)}\left[ \frac{e^{-\beta \Omega^{(i)}}}{\mathcal{Z}^{(i)}}- \frac{e^{-\beta{\omega}}+e^{-\beta\Omega^{(i-1)}}}{2\mathcal{Z}^{(i-1)}}\right].
\label{worki}
\end{aligned}
\right.
\ee
To guarantee the positive work extraction during each round of the protocol, a chain of constraints
\begin{equation}
 \tilde{\omega}^{(i)} < \frac{1}{\beta} \log \left[ \frac{2 + e^{-\beta \omega} + e^{-\beta \Omega^{(i-1)}}}{(1 + e^{-\beta \omega})(1 + e^{-\beta \tilde{\omega}^{(i-1)}})} \right] < \tilde{\omega}^{(i-1)},
\label{constrchain}
\end{equation}
should be satisfied, which means asymptotically $\displaystyle\lim_{n\rar\infty} \tilde{\omega}^{(n)}\rar0$. This implies that the amount of work extracted in subsequent cycles decreases progressively until it approaches zero when the system relaxes to the pure Gibbs state $\gamma_S $. As shown by Eq.\eqref{thermalstate}, the system at that stage has no further quantum coherence that can be utilized to trade for work extraction.

Finally, we need to identify the conditions under which the maximum total work extraction can be achieved after completing all protocol rounds. Since the protocols are conducted sequentially, we only need to determine the optimal $\tilde{\omega}_{\text{optimal}}^{(i)}$ for each cycle. From \eqref{worki}, we solve $d W^{(i)}_{\text{protocol-1}}/d\tilde{\omega}^{(i)}=0$, which gives a recurrence relation for $\tilde{\omega}_{\text{optimal}}^{(i)}$ ($i\geqslant2$) as:
\be
\left[\mathcal{Z}^{(i)}\right]^2-\frac{\mathcal{Z}^{(i)}}{e^{\beta\Omega^{(i)}} }\mathcal{R}^{(i-1)}+ \frac{\tilde{\omega}^{(i)}}{e^{\beta\Omega^{(i)}}}\beta\left(1+e^{-\beta\omega}\right)\mathcal{R}^{(i-1)}=0,
\ee 
where the function $\mathcal{R}^{(i-1)}$ determined by $(i-1)$-th cycle:
\be
\mathcal{R}^{(i-1)}:=2\left(1+\frac{e^{\beta\omega}}{1+e^{-\beta\tilde{\omega}_{\text{optimal}}^{(i-1)}}}\right) 
\ee 
and the initial condition of the relation is $\tilde{\omega}^{(1)}_{\text{optimal}}$ given by \eqref{optimal}.


In summary, after preparing the asymptotic steady state of a degenerate three-level \textsf{V} system with noise-induced quantum coherence, Protocol-1 provides a bath-assisted method to extract work from quantum coherence "fuel". Repeatly performing the Protocol-1 can fully convert the remaining quantum coherence into practical work $W_{\text{protocol-1}}:=\sum_iW_{\text{protocol-1}}^{(i)}$, eventually driving the \textsf{V} system to asymptotically approach a Gibbs state $\gamma_S$.

Nevertheless, it is worth emphasizing that Protocol-1 still exhibits subtle shortcomings. First, the protocol needs to be repeated enough times to achieve the optimal work extraction $W_{\text{protocol-1}}$. As illustrated above, to reach the final Gibbs state $\gamma_S$ when the process stops, we need to perform $n\rar\infty$ rounds to ensure $\tilde{\omega}^{(n)}\rar0$. On the other hand, we observe that the optimal work extraction in the protocol does not automatically ensure the MEW. In particular, we calculate the FED by substituting state \eqref{eq.34} into Eq.\eqref{work} and obtain
\be
 \Delta F_\beta(\rho_0)=\frac{1}{\beta} \log \left(\frac{1+2 {e}^{-\beta \omega}}{1+{e}^{-\beta \omega}}\right)
\ee

\begin{figure}[hbtp]
\centering  
\includegraphics[width=0.45\textwidth]{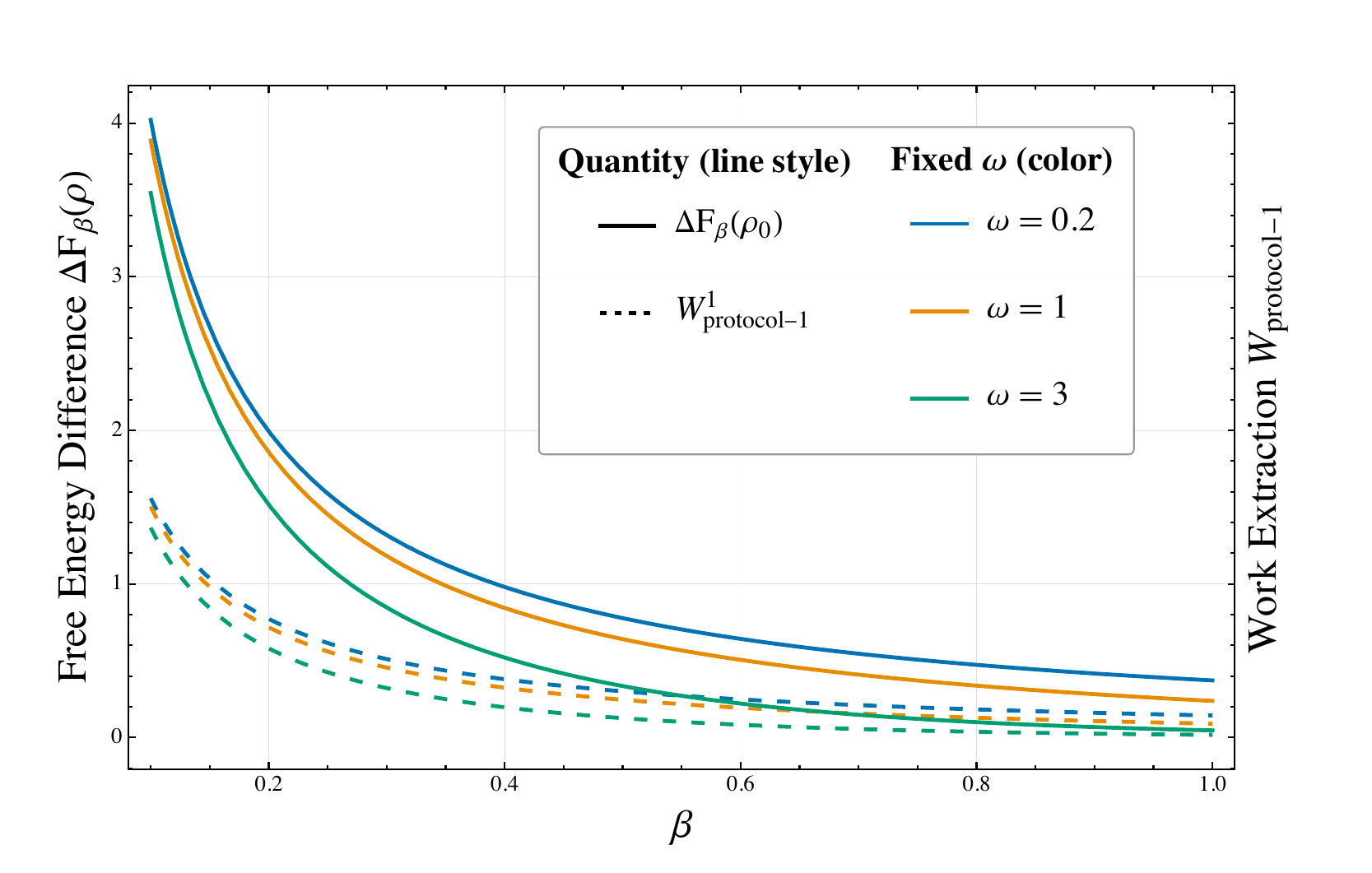}
\caption{{Comparison of the maximal work extraction $W_{\text {protocol-1}}^{(1)}$ (dashed lines) from the first round of Protocol-1 with the FED $\Delta F_\beta\left(\rho_0\right)$ (solid lines) between the initial state and the Gibbs state, as a function of the inverse bath temperature $\beta$. The estimation is performed for different atomic level spacings, i.e., $\omega=0.2$ (blue), $\omega=1$ (orange), and $\omega=3$ (green), respectively. }} \label{W-FED-beta}
\end{figure}

\begin{figure}[hbtp]
\centering  
\includegraphics[width=0.45\textwidth]{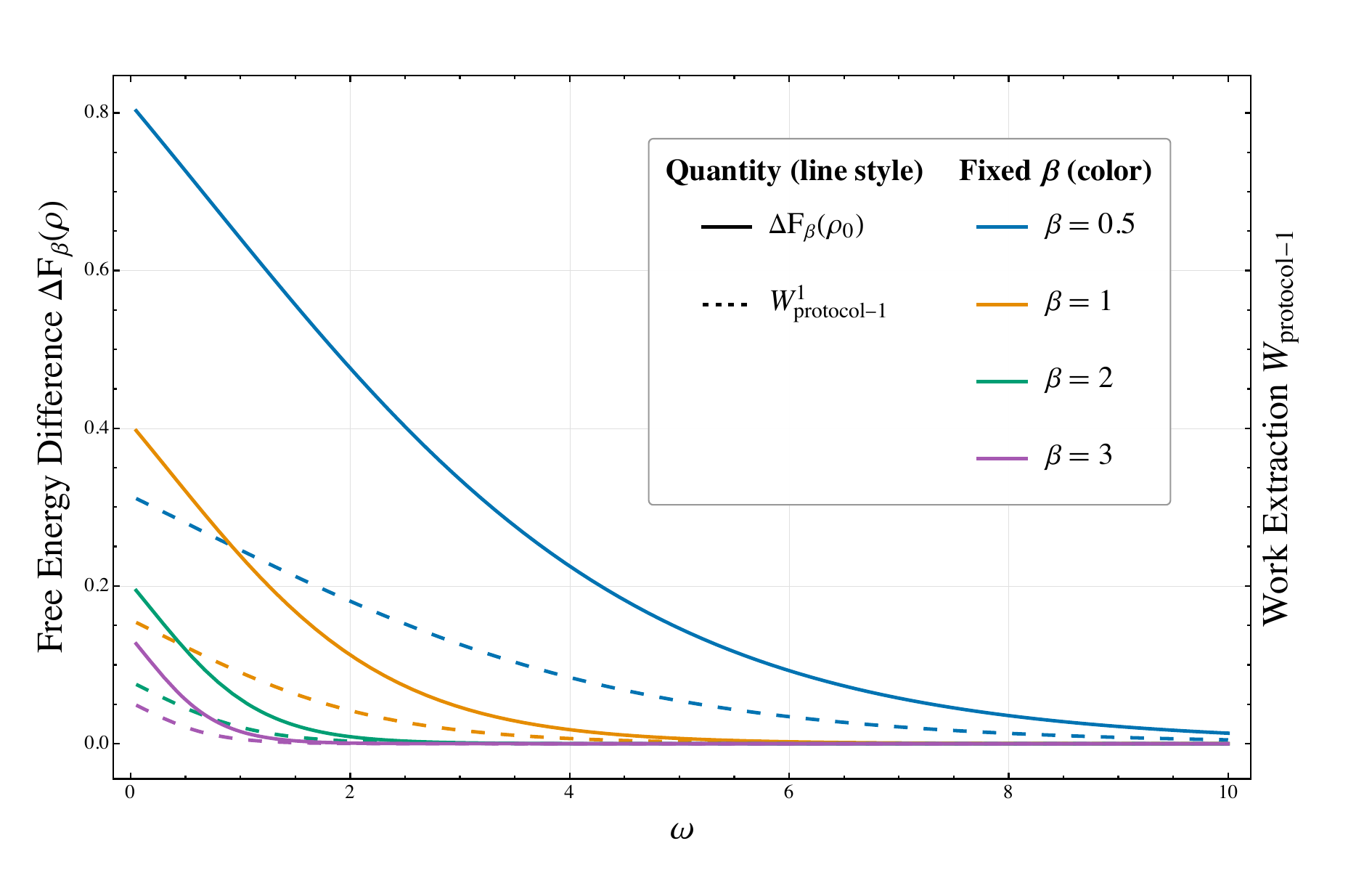}
\caption{{Comparison of the maximal work extraction $W_{\text {protocol-1}}^{(1)}$ (dashed lines) from the first round of Protocol-1 with the FED $\Delta F_\beta\left(\rho_0\right)$ (solid lines) between the initial state and the Gibbs state, as a function of the atomic level spacing $\omega$. The estimation is performed for different bath temperatures, i.e., $\beta=0.5$ (blue), $\beta=1$ (orange), $\beta=2$ (green), and $\beta=3$ (purple), respectively.}} \label{W-FED-omega}
\end{figure}

{In Fig. \ref{W-FED-beta} and Fig. \ref{W-FED-omega}, we numerically compare this FED to the value of optimal work $W_{\text{protocol-1}}$. It is evident that the optimal work $W_{\text{protocol-1}}$ extracted from the first round of Protocol-1 is generally less than the FED between the initial state $\rho_0$ of the protocol and the final $\gamma_S$, confirming that multiple rounds are required to approach the maximum extractable work. From the perspective of resource conversion, this indicates that Protocol-1 is not fully efficient, as the extracted work is smaller than the MEW of the three-level system. On the other hand, we also note that both protocol-extracted work and FED monotonically decrease as $\beta$ increases (see Fig. \ref{W-FED-beta}) or as the level spacing $\omega$ increases (see Fig. \ref{W-FED-omega}). This indicates that less work can be extracted from a thermal bath at lower temperatures, while a larger energy gap suppresses thermal excitation and reduces the free energy stored in the initial coherence.}

\subsection{Single-shot work extraction protocol}
\label{protocol2}

To address the shortcomings of Protocol-1, we propose a single-shot protocol, referred to as \textbf{Protocol-2}, using an idealized thermodynamic cycle in which the quantum coherence of the system is exhausted, and the MEW is fully achieved in a single process.

Preparing the degenerate three-level system with Hamiltonian \eqref{eq.1} in a state in the subspace spanned by $\{ \rho_{21}, \rho_{12}, \rho_{22}, \rho_{11} \}$. For later convenience, it could be rewritten as 
\begin{equation}
\rho_{S} =\frac{1}{2} \left(\begin{array}{cc}
\displaystyle {(1-b)\left( \mathbb{I}_{2\times 2}+\bm{n}\cdot\bm{\sigma}\right)} & 0 \\
0 & 2b
\end{array}\right)
\label{eq.48}
\end{equation}
where ground-state population $b=\rho_{00}$, $\bm{\sigma}$ denotes the vector of Pauli matrices, and $\bm{n}=|\bm{n}|\left(\sin\theta\cos\varphi,\sin\theta\sin\varphi,\cos\theta\right)$. 

{As the initial state for Protocol-2, $\rho_S$ is more general than what we used in Protocol-1. That is, by taking $b=(1+e^{-\beta \omega})^{-1}$ and a unit vector with $\theta=\pi/2,~\varphi=0$, we recover \eqref{eq.34} the initial state of Protocol-1 from the state \eqref{eq.48}.} The eigenvalues of $\rho_S$ are calculated as
\be
\lambda_0=b,~~~~\lambda_\pm=\frac{(1-b)(1\pm|\bm{n}|)}{2},
\ee
for future use.

\begin{figure}[hbtp]
\centering  
\includegraphics[width=0.48\textwidth]{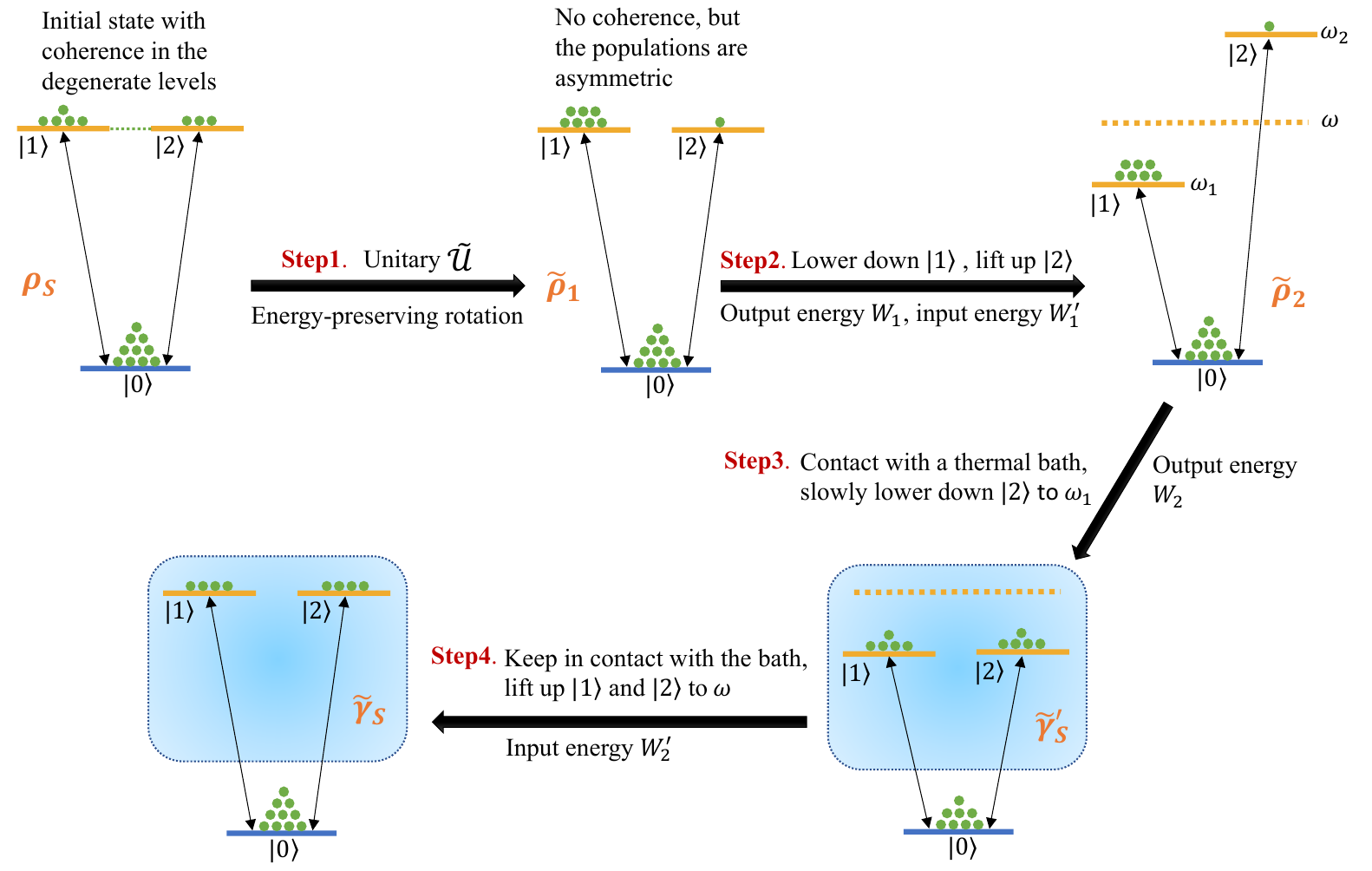}
\caption{Schematic of the work extraction Protocol-2 from a degenerate three-level quantum system. Starting from a coherence-containing initial state, the protocol involves an energy-preserving rotation, population reshuffling by lifting/lowering excited levels in isolation, then coupling to an ancillary thermal bath while adiabatically lowering one excited level, and finally restoring the energy structure in thermal contact. In the quasistatic limit, the protocol extracts the full free energy difference.}
\label{Protocol Long}
\end{figure}

Protocol 2 makes two changes relative to Protocol 1. First, in contrast to adiabatically lowering the level $\ket{2}$ of the three-level system in isolation, we consider lowering $\ket{2}$ adiabatically in Protocol-2 \textit{while} the system remains in contact with the thermal bath. Second, both excited levels are assumed to be dynamically adjusted. The scheme of Protocol-2 is illustrated in Fig.\ref{Protocol Long}.

We outline the steps of the work-extraction \textbf{Protocol-2} as follow:\\

\noindent\textbf{Step 1}: The cycle starts from $\rho_S$ \eqref{eq.48}, which has nonvanishing quantum coherence. Using an energy-conserving unitary:
\be
\widetilde{\mathcal{U}}=\left(\begin{array}{ccc}
-\sin \frac{\theta}{2} & e^{-i \varphi} \cos \frac{\theta}{2} & 0 \\
e^{i \varphi} \cos \frac{\theta}{2} & \sin \frac{\theta}{2} & 0 \\
0 & 0 & 1
\end{array}\right)
\ee
which satisfies $[\widetilde{\mathcal{U}},H_S]=0$, the initial state is rotated into $\widetilde{\mathcal{U}}\rho_S\widetilde{\mathcal{U}}^\dagger$ as:
 \be
 \widetilde{\rho}_1=\frac{1}{2}\left(\begin{array}{ccc}
(1-b)(1-|\bm{n}|) & 0 & 0 \\
0 & (1-b)(1+|\bm{n}|) & 0 \\
0 & 0 & 2b
\end{array}\right), 
 \ee
 which is diagonal with an asymmetry in population between the excited levels.
 
\noindent\textbf{Step 2}: The energy of $\ket{1}$ is lowered from $\omega$ to $\omega_1$, and the energy of $\ket{2}$ is raised from $\omega$ to $\omega_2$ in isolation. We require that $\omega_1$ and $\omega_2$ be fine-tuned to adjust the final state after completing the cycle, which is expected to be a Gibbs state similar to \eqref{thermalstate}. In particular, we need the energy shifts $\omega_1$ and $\omega_2$ to match the populations:
 \begin{equation}
\left\{
\begin{aligned}
\frac{e^{-\beta \omega_1}}{1+e^{-\beta \omega_1}+e^{-\beta \omega_2}} &= \frac{(1-b)(1+|\bm{n}|)}{2}, \\
\frac{e^{-\beta \omega_2}}{1+e^{-\beta \omega_1}+e^{-\beta \omega_2}} &= \frac{(1-b)(1-|\bm{n}|)}{2}.
\end{aligned}
\right.
\label{matching}
\end{equation}
This guarantees that after Step 2, the state of the three-level system can be characterized by energy levels like:  
 \be
\widetilde{\rho}_2=\frac{1}{\mathcal{Z}_{S}} \left(\begin{array}{ccc}
{e^{-\beta \omega_2}} & 0 & 0 \\
0 & {e^{-\beta \omega_1}}& 0 \\
0 & 0 &  {1}
\end{array}\right)
 \ee
with $\mathcal{Z}_{S}:=1+e^{-\beta \omega_1}+e^{-\beta \omega_2}$ is defined.

The output work from lowering $\ket{1}$ (i.e., $W_1$) and the input energy for raising $\ket{2}$ (i.e., $W'_1$) are: 
\begin{equation}
\begin{aligned}
W_1 &= \frac{1}{2}(1-b)(1+|\bm{n}|)(\omega - \omega_1), \\
W'_1 &= \frac{1}{2}(1-b)(1-|\bm{n}|)(\omega_2 - \omega).
\end{aligned}
\end{equation}

\noindent\textbf{Step 3}: The level $\ket{2}$ is slowly lowered from $\omega_2$ to $\omega_1$. Unlike in Protocol-1, during the level shifting, the system remains weakly coupled to the ancillary thermal bath at inverse temperature $\beta$. During the quasistatic process, the system state evolves through a continuous family of Gibbs states and arrives at
\be
\widetilde{\gamma}'_S=\frac{1}{1+2e^{-\beta\omega_1}} \left(\begin{array}{ccc}
{e^{-\beta \omega_1}} & 0 & 0 \\
0 & {e^{-\beta \omega_1}}& 0 \\
0 & 0 &  {1}
\end{array}\right)
\ee 
The resulting work extraction is
\begin{equation}
W_2 = \int_{\omega_2}^{\omega_1} \frac{-e^{-\beta {\omega'}} \, d {\omega'}}{1+e^{-\beta \omega_1}+e^{-\beta {\omega'}}} = \frac{1}{\beta} \ln \left(\frac{1+2 e^{-\beta \omega_1}}{1+e^{-\beta \omega_1}+e^{-\beta \omega_2}}\right).
\end{equation}

\noindent\textbf{Step 4}: Finally, both $\ket{1}$ and $\ket{2}$ are lifted back from $\omega_1$ to $\omega$ quasistatically, while maintaining thermal contact. This step leads the system to the final Gibbs state as
\be
\widetilde{\gamma}_S=\frac{1}{1+2e^{-\beta\omega}} \left(\begin{array}{ccc}
{e^{-\beta \omega}} & 0 & 0 \\
0 & {e^{-\beta \omega}}& 0 \\
0 & 0 &  {1}
\end{array}\right)
\label{final-2}
\ee 
and consumes energy:
\begin{equation}
W'_2 = \int_{\omega_1}^{\omega} \frac{2e^{-\beta {\omega'}} \, d {\omega'}}{1+2e^{-\beta {\omega'}}} = -\frac{1}{\beta} \ln \left(\frac{1+2 e^{-\beta \omega}}{1+2e^{-\beta \omega_1}}\right).
\end{equation}

The total extracted work after the cycle is $W_{\text{protocol-2}}= W_1 + W_2 - W'_1 - W'_2$ can explicitly be given
\bea
W_{\text{protocol-2}} 
&=&\omega(1-b)+\frac{1}{\beta}\log \left(\frac{1+2 e^{-\beta \omega}}{1+e^{-\beta \omega_1}+e^{-\beta \omega_2}}\right)\no\\
&&-\frac{(1-b)}{2}\left[(1+|\bm{n}|) \omega_1+(1-|\bm{n}|) \omega_2\right]~~~~
\eea
{Under the matching condition \eqref{matching}, the extracted work $W_{\text{protocol-2}}$ is independent of the energy shifts $\omega_1$ and $\omega_2$, and can be confirmed to be exactly the FED
\bea
 \Delta F_\beta(\rho_S)&=&\omega(1-b)+\frac{1}{\beta}\bigg[\log\left(1+2e^{-\beta \omega}\right)\no\\
&&\left.-H_{\text{bin}}(b)-(1-b)H_{\text{bin}}\left(\frac{1+|\bm{n}|}{2}\right)\right]~~~~~~
\label{FED-p2}
\eea
between the initial state $\rho_S$ of the three-level system and its final thermal state $\widetilde{\gamma}_S$ after the cycle is completed.} 

\begin{figure}[hbtp]
\centering
\includegraphics[width=0.5\textwidth]{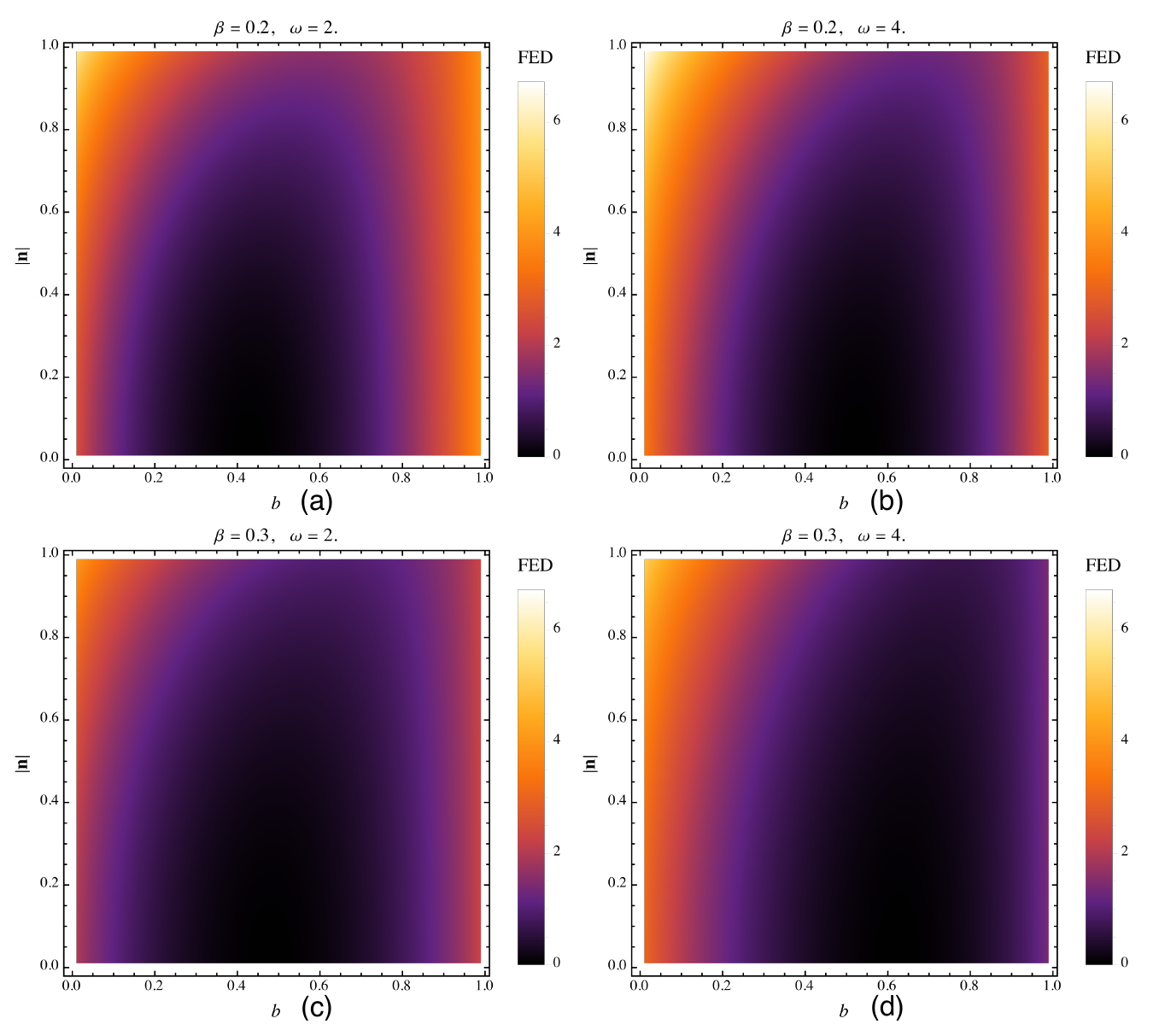}
\caption{{The FED as a function of $b=\rho_{00}$ and $|\bm{n}|$. The four panels set the state as in (a) $(\beta,\omega)=(0.2,2)$, in (b) $(\beta,\omega)=(0.2,4)$, in (c) $(\beta,\omega)=(0.3,2)$, and in (d) $(\beta,\omega)=(0.3,4)$, respectively. 
}}
\label{generalFEDmap-2}
\end{figure}

{In Fig. \ref{generalFEDmap-2}, we map the FED \eqref{FED-p2} over the state parameters $b=\rho_{00}$ and $|\boldsymbol{n}|$, for different settings of ancillary bath temperature and energy spacing. One observes that the FED increases monotonically with $|\boldsymbol{n}|$ and approaches its maximum with maximal coherence ($|\boldsymbol{n}|=1$). However, the FED changes non-monotonically with growing population $b$. For $|\boldsymbol{n}|=0$, the FED is vanishing when $b_{\text{eq}}:=(1+e^{-\beta \omega})^{-1}$, corresponding to a Gibbs state (i.e., $\rho_S\rar\widetilde{\gamma}_S$). For $|\boldsymbol{n}|\neq 0$, one can confirm that the minimum of FED shifts away to the right side of $b_{\text{eq}}$. }

In summary, the single-shot Protocol-2 saturates the upper bound imposed by FED and extracts the MEW of the three-level system. As we have presented, the key resource is the quantum coherence that is exhausted after Step 1, resulting in an asymmetry in the excited-state populations. 

\section{Summary and discussion}
\label{sec5}

In this work, we analyze a (nearly) degenerate three-level \textsf{V}-type system coupled to a thermal bath via dipole-monopole interactions, focusing on the emergence and utility of its quantum coherence for work extraction. We demonstrate that long-lived coherence can persist in asymptotic steady states when the degenerate transition dipoles are aligned, even under continuous thermalization. To exploit this resource, we propose two protocols that extract practical work via an energy-preserving unitary operation, converting coherence into a population asymmetry between the degenerate states. This asymmetry enables work extraction through a designed thermodynamic cycle that uses an ancillary bath and controlled level shifts. In particular, Protocol-1 is repeatable, consuming coherence incrementally in each cycle, whereas Protocol-2 is a single-shot thermal process that extracts practical work equal to the FED of the initial state, a fundamental bound on the MEW in contact with a thermal reservoir.


The concrete protocols designed for a degenerate three-level  \textsf{V} quantum system illustrate how long-lived quantum coherence can serve as a resource that can be converted into practical work. The same analysis can be extended to the nearly degenerate system, with more involved dynamics in \ref{C4}. Besides deepening understanding of the thermodynamic role of quantum coherence, our analysis also opens the possibility of utilizing other quantum resources, such as quantum discord \cite{Gu2016}. The protocols also provide a basis for designing future coherence-driven quantum thermal machines \cite{Binder2019} or quantum devices/batteries \cite{Campaioli2024} using a three-level quantum system.

\begin{acknowledgments}
This work is supported by the National Natural Science Foundation of China (No. 12475061), Natural Science Basic Research Plan in Shaanxi Province of China (No. 2026JC-QYCX-002), and the Innovation Program for Quantum Science and Technology (2021ZD0302400). J.F. would like to thank Xueyuan Hu for an insightful discussion.
\end{acknowledgments}

\appendix

\section{Gell-Mann matrices}
\label{appA}

Together with the identity matrix, the Gell-Mann matrices $\{ \lambda_i \}_{i=1}^{8}$ form a complete traceless Hermitian basis for $3 \times 3$ matrices, 
satisfying the normalization $\operatorname{Tr}(\lambda_i \lambda_j) = 2 \delta_{ij}$. For added convenience, we introduce an alternative basis $\{ \mathbb{P}_i\}_{i=1}^8$ constructed by combinations of Gell-Mann matrices:
\begin{equation}
\begin{aligned}
\mathbb{P}_1 &= T_{+} = \frac{ \lambda_1 + i \lambda_2 }{2} \quad , \quad \mathbb{P}_2 = T_{-} = \frac{ \lambda_1 - i \lambda_2 }{2}, \\
\mathbb{P}_3 &=V_{+} = \frac{\lambda_4 + i \lambda_5 }{2} \quad , \quad \mathbb{P}_4= V_{-} = \frac{\lambda_4 - i \lambda_5 }{2},\\
\mathbb{P}_5 &=U_{+} = \frac{\lambda_6 + i \lambda_7 }{2} \quad,\quad \mathbb{P}_6 =U_{-} = \frac{\lambda_6 - i \lambda_7 }{2} , \\
\mathbb{P}_7 &=V_3 = \frac{\sqrt{3} \lambda_8 + \lambda_3}{2} ~, ~~~
\mathbb{P}_8=U_3 = \frac{\sqrt{3} \lambda_8 - \lambda_3}{2}.
\end{aligned}
\end{equation}
Denote $\mathbb{P}_0:=\mathbb{I}_{3 \times 3}$, then an arbitrary $3 \times 3$ matrix $A$ can be expanded as $A=\sum_{i=0}^8 a_i\mathbb{P}_{i}$.

\section{Lindblad dissipator for a nearly degenerate case} \label{Appendix A}

Assume the dipole moments of this \textsf{V} system with unit magnitude, i.e., $|\mathbf{d}_{(i)}| = 1$ for $i=1,2$, and the dipole alignment parameter as $\mathfrak{p} = \cos{\Theta_{12}}$, where $\Theta_{12}$ is the angle between $\mathbf{d}_{(1)}$ and $\mathbf{d}_{(2)}$. For a small energy splitting $\delta = \omega_2 - \omega_1$ between the excited states, we retain the terms oscillating as $e^{\pm i \delta t}$ in the derivation of the master equation. Following the standard microscopic derivation of quantum master equation \cite{Open1}, the corresponding superoperator in this case is given by
\bea
\mathcal{L}_2[\rho] &=& \sum_{i=1}^2 \sum_{\pm} \Gamma^{\pm}_{(i)} \mathcal{D}\left(\sigma_{\mp}^{(i)}, \sigma_{\pm}^{(i)}\right) 
+ \frac{\mathfrak{p}}{2} \sum_{k=1}^2 \sum_{k' \neq k} \sum_{\pm} \Gamma^{\pm}_{(k)}\no\\
&&\times \left[ e^{\pm i (k' - k) \Delta t} \mathcal{Q}\left(\sigma_{\mp}^{(k)}, \sigma_{\pm}^{(k')}\right) \right.\no\\
 &&+\left. e^{\pm i (k - k') \Delta t} \mathcal{P}\left(\sigma_{\mp}^{(k')}, \sigma_{\pm}^{(k)}\right) \right],
 \label{A1}
 \eea
where $\mathcal{D}(a, b) = a \rho b - \frac{1}{2} \{ b a, \rho \}$ is the standard dissipator, and $\mathcal{Q}(a, b) = a \rho b - b a \rho$, $\mathcal{P}(a, b) = a \rho b - \rho b a$ are generalized dissipators capturing the non-secular terms.

Due to the energy splitting between the excited states, the validity of the superoperator \eqref{A1} is restricted within timescale $\tau_S \ll t \ll 1/\delta$. For very late times $t \gg 1/\delta$, or in the steady-state limit $t \to \infty$, all the fast oscillating terms (including all pathways for transitions $\ket{1} \leftrightarrow \ket{0}$, $\ket{2} \leftrightarrow \ket{0}$, and $\ket{1} \leftrightarrow \ket{2}$) in the master equation would average out by the RWA assumption. As a result, the Lindblad dissipator \eqref{A1} reduces to
\begin{equation}
\mathcal{L}_2[\rho] \longrightarrow \sum_{i=1}^2 \sum_{\pm} \Gamma^{\pm}_{(i)} \mathcal{D}\left(\sigma_{\mp}^{(i)}, \sigma_{\pm}^{(i)}\right),
\label{B2}
\end{equation}
indicating that the three-level system behaves like two independent two-level open quantum systems, which approach Gibbs states undergoing thermalization. On the other hand, in the limit of a degenerate scenario, i.e., $\delta\rar0$, one can recover the result of Eq.\eqref{eq1.4} as expected.

\section{Perturbative dynamics of a nearly degenerate case} 
\label{appendixc}

To quantify the influence of the tiny splitting $\delta$ on the dynamics, we employ a perturbative expansion:
\begin{equation}
\widetilde{\bm{\Pi}}(t) \approx \bm{\Pi}(t) + \delta \bm{\Pi}_1(t), \qquad \widetilde{\mathbf{M}} \approx \mathbf{M} + \delta \mathbf{M}_1.
\label{C1}
\end{equation}
Substituting these expansions into the Eq.\eqref{eq.nd} and retaining terms up to first order in $\delta$, gives
\begin{subequations}
\begin{eqnarray}
&&\dot{\bm{\Pi}}  = \mathbf{M}\cdot\bm{\Pi} - \mathbf{b}, \label{C2a}\\
&&\dot{ \bm{\Pi}}_1 =\mathbf{M}\cdot  \bm{\Pi}_1 + \mathbf{M}_1 \cdot \bm{\Pi}(t).\label{C2b}
\end{eqnarray} \label{eq.C2}
\end{subequations}
Note that Eq.\eqref{C2a} is satisfied by $\bm{\Pi}(t)$ and $\mathbf{M}$ and is precisely Eq.\eqref{eq.16}, confirming their role defined for the degenerate case. On the other hand, the leading-order perturbations $\bm{\Pi}_1$ and $\mathbf{M}_1$ satisfy Eq.\eqref{C2b}, while the explicit form of $\mathbf{M}_1$ can be separated from Eq.\eqref{eqMnd} as 
\begin{equation}
\mathbf{M}_1 =
\left(\begin{array}{cccc}
-\gamma_{+}' & \gamma_{-}' & 0 & 0 \\
\gamma_{+}' & -\gamma_{-}' & \mathfrak{p} \gamma_{+}' & 0 \\
-\displaystyle \frac{1}{2} \mathfrak{p} \gamma_{+}' & \displaystyle\frac{1}{2} \mathfrak{p} \gamma_{-}' & \displaystyle-\frac{1}{2} \gamma_{+}' & -i \\
0 & 0 & -i &\displaystyle -\frac{1}{2} \gamma_{+}'
\end{array}\right),
\end{equation}
with the shorthand for differential rates as $\gamma_{\pm}' =\left(\gamma_{\pm,(2)} - \gamma_{\pm,(1)}\right) / \delta$ has been used.

From Eq.\eqref{C2b}, we see that once the degenerate solution ${\bm{\Pi}}$ is obtained, it serves as a time-dependent source term for the leading-order correction $\bm{\Pi}_1(t)$. In the aligned dipoles ($\mathfrak{p}=1$) scenario, we set the same initial conditions as in Sec.\ref{II.2}. Then, the first-order inhomogeneous equation Eq.\eqref{C2b} can be solved, with the source $\bm{\Pi}(t)$ given by Eq.\eqref{eqq} and initial condition $\bm{\Pi}(0)=\bm{0}$, which yields the components of the leading-order correction $\bm{\Pi}_1(t)$ as
\begin{widetext}
\begin{equation}
\left\{
\begin{aligned}
(\rho_{22})_1 &= e^{-2(1 + x)\gamma_{+,(1)} t}
\left[\frac{b_1}{4(1 + x)} + \frac{b_2 - b_3 - 2a_3}{2(1 + 2x)} - \frac{b_3}{2} \gamma_{+,(1)} t - \frac{d}{2(1 + x)(1 + 2x)\gamma_{+,(1)}}\right]
+ a_1  \\
&~~~~+ e^{-\gamma_{+,(1)} t}
\left[\frac{2a_3 - b_2 + b_3 + 2d/\gamma_{+,(1)}}{2(1 + 2x)} + \frac{b_2 \gamma_{+,(1)} t - b_1}{2} + a_2 \gamma_{+,(1)} t - a_1\right]+ \frac{b_1\gamma_{+,(1)}(1 + 2x) - 2d}{4(1 + x)\gamma_{+,(1)}}, \\[1ex]
(\rho_{00})_1 &= e^{-2(1 + x)\gamma_{+,(1)} t}
\left[\frac{d}{(1 + x)(1 + 2x)\gamma_{+,(1)}} - \frac{b_1}{2(1 + x)} - \frac{b_2}{(1 + 2x)} + b_3 \gamma_{+,(1)} t\right] \\
&~~~~+ e^{-\gamma_{+,(1)} t} \frac{b_2\gamma_{+,(1)} - 2d}{(1 + 2x)\gamma_{+,(1)}} 
+ \frac{b_1\gamma_{+,(1)}+2d }{2(1 + x)\gamma_{+,(1)}}, \\[1ex]
(\rho_{+})_1 &= e^{-2(1 + x)\gamma_{+,(1)} t}
\left[\frac{b_1}{4(1 + x)} + \frac{b_2}{2(1 + 2x)} - \frac{b_3 \gamma_{+,(1)} t}{2} - \frac{d}{2(1 + x)(1 + 2x)\gamma_{+,(1)}}\right] \\
&~~~~~ - e^{-\gamma_{+,(1)} t} \frac{b_2\gamma_{+,(1)} + 4x d}{2(1 + 2x)\gamma_{+,(1)}} 
+ \frac{(1 + 2x)d}{2(1 + x)\gamma_{+,(1)}} - \frac{b_1}{4(1 + x)}, \\[1ex]
(\rho_{-})_1 &= i \left[ \frac{d}{2(1 + 2x)} + e^{-\gamma_{+,(1)} t} \left(\frac{{(1+2 x)(b+2 c)-1}}{4\gamma_{+,(1)}(1+x)} - \frac{d \gamma_{+,(1)} t}{2} - \frac{d}{2(1 + 2x)} \right) - \frac{{(1+2 x)(b+2 c)-1}}{4\gamma_{+,(1)}(1+x)} \right],
\end{aligned}
\right.\label{C4}
\end{equation}
\end{widetext}
where the coefficients $\{a_1,a_2,a_3,b_1,b_2,b_3\}$ are the combinations of the parameters $\{a,b,c\}$ specifying the system's initial state:
\begin{equation}
\left\{
\begin{aligned}
a_1 &= \frac{2\gamma_{-}' (1 + b + 2c) - \gamma_{+}' (1 + 2x - b - 2c)}{4\gamma_{+,(1)}(1 + x)}, \\
a_2 &= \frac{\gamma_{+}' (1-2a - b )}{2\gamma_{+,(1)}}, \\
a_3 &= \frac{2\gamma_{-}'+ \gamma_{+}' }{\gamma_{+,(1)}}\frac{(1 + 2x)b-1 - 2c }{4(1 + x)},
\end{aligned}
\right.
\end{equation}
\begin{equation}
\left\{
\begin{aligned}
b_1 &= \frac{(\gamma_{+}' - \gamma_{-}' ) (1+b+2 c)}{2\gamma_{+,(1)}(1+x)},\\
b_2 &= \frac{\gamma_{+}' (2 a+b-1)}{2\gamma_{+,(1)}}, \\
b_3 &= \frac{(\gamma_{+}' + \gamma_{-}' )[1+2 c-(1+2 x) b]}{2\gamma_{+,(1)}(1+x)}.
\end{aligned}
\right.
\end{equation}
The full density matrix of a nearly degenerate three-level system is then determined by substituting \eqref{C4} back into the coherent vector \eqref{C1}.

\section{Final state of work-extraction protocol} 
\label{AppendixD}

Step 5 of the Protocol-1 proposed in Section \ref{protocol1} consists of a thermalization of system state $\rho_1^{(1)}$. Since the system is again degenerate after the energy level of $\ket{2}$ is adjusted from $\Omega^{(1)}$ to $\omega$, we can choose $\rho_2^{(1)}$ as the initial state of the solution Eq.(\ref{eq.20}). The parameters can be read as 
\be
a=\frac{e^{-\beta\Omega^{(1)}}}{\mathcal{Z}^{(1)}},~~b=\frac{1}{\mathcal{Z}^{(1)}},~~c=d=0
\ee
Substituting them into Eq.(\ref{eq.20}), we obtain the density matrix after completing Protocol-1:
\begin{equation}
\rho_{\text{final}}^{(1)} =
\left(\begin{array}{ccc}
\displaystyle \frac{ \mathcal{K}^{(1)}}{\mathcal{M}^{(1)}} &
\displaystyle\frac{1 - e^{-\beta \tilde{\omega}^{(1)}}}{\mathcal{M}^{(1)}} & 0 \\
\displaystyle\frac{1 - e^{-\beta \tilde{\omega}^{(1)}}}{\mathcal{M}^{(1)}} &
\displaystyle\frac{ \mathcal{K}^{(1)}}{\mathcal{M}^{(1)}} & 0 \\
0 & 0 &\displaystyle \frac{2e^{-\beta \omega}(1 + \mathcal{Z}^{(1)})}{\mathcal{M}^{(1)}}
\end{array}\right),
\label{eq.D2}
\end{equation}
where abbreviation $ \mathcal{K}^{(1)}:=1 + e^{-\beta \tilde{\omega}^{(1)}} + 2 \mathcal{Z}^{(1)}$ and $\mathcal{M}^{(1)}:=4(1 + e^{\beta \omega})\mathcal{Z}^{(1)}$ are used.

Similarly, the density matrix after completing Protocol-$1^{(n)}$ is obtained by initializing Eq.(\ref{eq.20}) with the thermal state $\rho^{(n)}_{2}$. The solution, denoted by $\rho_{\text{final}}^{(n)}$, has the same form as $\rho_{\text{final}}^{(1)}$, with all instances of $\tilde{\omega}^{(1)}$ replaced by $\tilde{\omega}^{(n)}$. Under constraints like \eqref{constr} for energy shifts in each cycle, we know that asymptotically $\lim_{n\rar\infty} \tilde{\omega}^{(n)} \rightarrow 0$. Consequently, the system eventually approaches a Gibbs state:
\begin{equation}
\gamma_S = \frac{1}{1 + 2 e^{-\beta \omega}}\left(\begin{array}{ccc}
e^{-\beta \omega}& 0 & 0 \\ 0 & e^{-\beta \omega} & 0 \\ 0 & 0 & 1
\end{array}\right)
\label{thermalstate}
\end{equation}
From this state, no further quantum coherence can be utilized to trade for practical work.



\nocite{*}

\begin{thebibliography}{99}

\bibitem{Vinjanampathy2016}
S.~Vinjanampathy and J.~Anders, \textit{Quantum thermodynamics}, \href{http://dx.doi.org/10.1080/00107514.2016.1201896}{Contemp. Phys. \textbf{57}, 545 (2016)}.

\bibitem{Goold2016}
J. Goold, M. Huber, A. Riera, L. del Rio, and P. Skrzypczyk, \textit{The role of quantum information in thermodynamics—a topical review}, \href{http://dx.doi.org/10.1088/1751-8113/49/14/143001}{J. Phys. A: Math. Theor. \textbf{49}, 143001 (2016)}.




\bibitem{Tscherbul2014}
T.~V.~Tscherbul and P.~Brumer, \textit{Long-lived quantum coherence in \textsf{V}-type systems driven by incoherent light}, \href{https://journals.aps.org/prl/abstract/10.1103/PhysRevLett.113.113601}{Phys. Rev. Lett. \textbf{113}, 113601 (2014)}.

\bibitem{Koyu2018}
S.~Koyu and T.~V.~Tscherbul,
\textit{Long-lived quantum coherences in a \textsf{V}-type system strongly driven by a thermal environment}, \href{https://doi.org/10.1103/PhysRevA.98.023811}
{Phys. Rev. A \textbf{98}, 023811 (2018)}.

\bibitem{Dodin2018}
A.~Dodin, T.~V.~Tscherbul, and P.~Brumer, \textit{Quantum dynamics of incoherently driven \textsf{V}-type systems: Analytic solutions beyond the secular approximation}, \href{https://doi.org/10.1063/1.4954243}{J. Chem. Phys. \textbf{148}, 184304 (2018)}.

\bibitem{Koyu2021}
S.~Koyu, A.~Dodin, P.~Brumer, and T.~V.~Tscherbul,
\textit{Steady-state Fano coherences in a \textsf{V}-type system driven by polarized incoherent light}, \href{https://doi.org/10.1103/PhysRevResearch.3.013295}{Phys. Rev. Research \textbf{3}, 013295 (2021)}.





\bibitem{Cao2016}
D. Xu, C. Wang, Y. Zhao, and J. Cao, \textit{Polaron effects on the performance of light-harvesting systems: a quantum heat engine perspective}, \href{https://iopscience.iop.org/article/10.1088/1367-2630/18/2/023003}{New J. Phys. \textbf{18}, 023003 (2016)}.

\bibitem{Avisar2016}
D. Avisar and A. D. Wilson-Gordon, \textit{Thermal-light-induced dynamics: Coherence and revivals in \textsf{V}-type and molecular Jaynes-Cummings systems}, \href{https://doi.org/10.1103/PhysRevA.93.033843}{Phys. Rev. A \textbf{93}, 033843 (2016)}.

\bibitem{Segal2018}
M. Kilgour and D. Segal, \textit{Coherence and decoherence in quantum absorption refrigerators}, \href{https://doi.org/10.1103/PhysRevE.98.012117}{Phys.
Rev. E 98, 012117 (2018)}.


\bibitem{Cao2018}
K. E. Dorfman, D. Xu, and J. Cao, \textit{Efficiency at maximum power of a laser quantum heat engine enhanced by noise-induced coherence}, \href{https://doi.org/10.1103/PhysRevE.97.042120}{Phys. Rev. E \textbf{97}, 042120 (2018)}.

\bibitem{Brumer2018}
P. Brumer, \textit{Shedding (Incoherent) Light on Quantum Effects in Light-Induced Biological Processes}, \href{https://pubs.acs.org/doi/10.1021/acs.jpclett.8b00874}{J. Phys. Chem. Lett. \textbf{9}, 2946 (2018)}.

\bibitem{Cresser2021}
J. D. Cresser and J. Anders, \textit{Weak and Ultrastrong Coupling Limits of the Quantum Mean Force Gibbs State}, \href{https://doi.org/10.1103/PhysRevLett.127.250601}{Phys. Rev. Lett. 127, 250601 (2021)}.




\bibitem{Streltsov2015}
A. Streltsov, U. Singh, H. S. Dhar, M. N. Bera, and G. Adesso, \textit{Measuring Quantum Coherence with Entanglement}, \href{http://dx.doi.org/10.1103/PhysRevLett.115.020403}{Phys. Rev. Lett. \textbf{115}, 020403 (2015)}.


\bibitem{Chitambar2016}
E. Chitambar and M.-H. Hsieh, \textit{Relating the Resource Theories of Entanglement and Quantum Coherence}, \href{https://doi.org/10.1103/PhysRevLett.117.020402}{Phys. Rev. Lett. \textbf{117}, 020402 (2016)}.


\bibitem{Gu2016}
J. Ma, B. Yadin, D. Girolami, V. Vedral, and M. Gu, \textit{Converting Coherence to Quantum Correlations}, \href{http://dx.doi.org/10.1103/PhysRevLett.116.160407}{Phys. Rev. Lett. \textbf{116}, 160407 (2016)}.



\bibitem{Plenio2017}
A. Streltsov, G. Adesso, and M. B. Plenio, \textit{Colloquium: Quantum coherence as a resource}, \href{https://doi.org/10.1103/RevModPhys.89.041003}{Rev. Mod. Phys. \textbf{89}, 041003 (2017)}.


\bibitem{Lostaglio2015}
M.~Lostaglio, D.~Jennings, and T.~Rudolph, \textit{Quantum coherence, time-translation symmetry, and thermodynamics}, \href{http://dx.doi.org/10.1103/PhysRevX.5.021001}{Phys. Rev. X \textbf{5}, 021001 (2015)}.



\bibitem{Campaioli2024}
{F. Campaioli, S. Gherardini, J. Q. Quach, M. Polini, and G. M. Andolina, \textit{Colloquium: quantum batteries},
\href{https://doi.org/10.1103/RevModPhys.96.031001}{Rev. Mod. Phys. \textbf{96}, 031001 (2024).}}


\bibitem{Ferraro2026}
{D. Ferraro, F. Cavaliere, M. G. Genoni, et al. \textit{Opportunities and challenges of quantum batteries},
\href{https://doi.org/10.1038/s42254-025-00906-5}{Nat. Rev. Phys. \textbf{8}, 115 (2026)}}


\bibitem{Quach2023}
{J.~Q.~Quach, G.~Cerullo, and T.~Virgili,
\textit{Quantum batteries: The future of energy storage?},
\href{https://doi.org/10.1016/j.joule.2023.09.003}{Joule \textbf{7}, 2195 (2023)}.}

\bibitem{Yang2023}
{X.~Yang, Y.-H.~Yang, M.~Alimuddin, R.~Salvia, S.-M.~Fei, L.-M.~Zhao, S.~Nimmrichter, and M.-X.~Luo,
\textit{Battery capacity of energy-storing quantum systems},
\href{https://doi.org/10.1103/PhysRevLett.131.030402}{Phys. Rev. Lett. \textbf{131}, 030402 (2023)}.}

\bibitem{Lai2024}
{P.-R.~Lai, J.-D.~Lin, Y.-T.~Huang, H.-C.~Jan, and Y.-N.~Chen,
\textit{Quick charging of a quantum battery with superposed trajectories},
\href{https://doi.org/10.1103/PhysRevResearch.6.023136}{Phys. Rev. Research \textbf{6}, 023136 (2024)}.}

\bibitem{Oularabi2025}
{A.~Oularabi, A.~El Allati, and K.~El Anouz,
\textit{Enhancing ergotropy of quantum batteries through coherence and non-Markovianity},
\href{https://doi.org/10.1016/j.physa.2025.131003}{Physica A \textbf{679}, 131003 (2025)}.}


\bibitem{Li2025}
{L. Li, S. Zhao, Y.-H. Shi, K. Xu, H. Fan, D. Zheng, and Z. Xiang,
\textit{Experimental extraction of coherent ergotropy and its energetic cost in a superconducting qubit},
\href{https://doi.org/10.1103/ptr1-59tv}{Phys. Rev. Research \textbf{7}, 043236 (2025)}.}

\bibitem{Niedenzu2015}
W. Niedenzu, D. Gelbwaser-Klimovsky, and G. Kurizki, \textit{Performance limits of multilevel and multipartite quantum heat machines}, \href{https://doi.org/10.1103/PhysRevE.92.042123}{Phys. Rev. E \textbf{92}, 042123 (2015)}.





\bibitem{Korzekwa2016}
K.~Korzekwa, M.~Lostaglio, J.~Oppenheim, and D.~Jennings, \textit{The extraction of work from quantum coherence}, \href{http://dx.doi.org/10.1088/1367-2630/18/2/023045}{New J. Phys. \textbf{18}, 023045 (2016)}.

\bibitem{Deffner2022}
N. M. Myers, O. Abah, and S. Deffner, \textit{Quantum thermodynamic devices: From theoretical proposals to experimental reality}, \href{https://doi.org/10.1116/5.0083192}{AVS Quantum Sci. \textbf{4}, 027101 (2022)}.

\bibitem{Scully}
M. S. Zubairy and M. O. Scully, \textit{Quantum Optics} (Cambridge University Press, 1997).


\bibitem{Open1}
H.-P. Breuer and F. Petruccione, \textit{The Theory of Open Quantum Systems} (Oxford University Press, 2002).

\bibitem{HaagHugenholtzWinnink1967}
R.~Haag, N.~M.~Hugenholtz, and M.~Winnink, 
\textit{On the Equilibrium States in Quantum Statistical Mechanics}, \href{https://doi.org/10.1007/BF01646342}{Commun. Math. Phys. \textbf{5}, 215 (1967)}.

\bibitem{Pusz1978}
W. Pusz and S. Woronowicz, \textit{Passive states and KMS states for general quantum systems}, \href{https://doi.org/10.1007/BF01614224}{Commun. Math. Phys. \textbf{58}, 273 (1978)}. 
\bibitem{Han2025}
S.-W. Han, W. Chen, L. Chen, Z. Ouyang, and J. Feng, 
\textit{Reliable quantum master equation of the Unruh–DeWitt detector}, 
\href{https://journals.aps.org/prd/abstract/10.1103/c9wz-lm5m}{Phys. Rev. D \textbf{112}, 085014 (2025).}



\bibitem{sec1-15}
G. Lindblad, On the generators of quantum dynamical semigroups, \href{https://doi.org/10.1007/BF01608499}{Commun. Math. Phys. 48, 119 (1976)}.

\bibitem{sec1-16}
V. Gorini, A. Kossakowski, and E. C. G. Sudarshan, Completely positive dynamical semigroups of N-level systems, \href{https://doi.org/10.1063/1.522979}{J. Math. Phys. 17, 821 (1976)}.





\bibitem{Vtype1}
D. Gelbwaser-Klimovsky, {et al.}, \textit{Power enhancement of heat engines via correlated thermalization in a three-level “working fluid”}, \href{https://doi.org/10.1038/srep14413}{Sci. Rep. \textbf{5}, 14413 (2015)}. 

\bibitem{Min2025}
B.~Min, M.~Gerry, and D.~Segal,
\textit{Separation of relaxation timescales via strong system–bath coupling: Dissipative three-level system as a case study},
\href{https://doi.org/10.1103/31cp-kq8f}{Phys. Rev. A \textbf{112}, 062226 (2025)}.


\bibitem{Sufiani2026}
R. Sufiani and S. Saei, 
\textit{Coherence freezing and tunable decoherence-free/uniformly-decaying subspaces in collectively decaying qutrits},
\href{https://doi.org/10.1016/j.physleta.2026.131348}{Phys. Lett. A \textbf{574}, 131348 (2026)}.


\bibitem{Alicki1987}
R. Alicki and K. Lendi, \textit{Quantum Dynamical Semigroups and Applications}, Lecture Notes in Physics Vol. 286 (Springer-Verlag, Berlin, 1987).


\bibitem{Lendi1987}
K. Lendi, \textit{Evolution matrix in a coherence vector formulation for quantum Markovian master equations of $N$-level systems}, \href{https://iopscience.iop.org/article/10.1088/0305-4470/20/1/011}{J. Phys. A \textbf{20}, 15 (1987)}.






\bibitem{Baumgratz_2014}
T. Baumgratz, M. Cramer, and M. Plenio, \textit{Quantifying Coherence}, \href{http://dx.doi.org/10.1103/PhysRevLett.113.140401}{Phys. Rev. Lett. \textbf{113}, 140401 (2014)}. 



%

\bibitem{Brandao2013}
F. Brand\~ao, M. Horodecki, J. Oppenheim, J. M. Renes, and R. W. Spekkens,
\textit{Resource theory of quantum states out of thermal equilibrium}, \href{https://journals.aps.org/prl/abstract/10.1103/PhysRevLett.111.250404}{Phys. Rev. Lett. \textbf{111}, 250404 (2013)}.


\bibitem{Allahverdyan2004}
{A.~E.~Allahverdyan, R.~Balian, and T.~M.~Nieuwenhuizen,
\textit{Maximal work extraction from finite quantum systems},
\href{https://doi.org/10.1209/epl/i2004-10101-2}{Europhys. Lett. \textbf{67}, 565 (2004)}.}

\bibitem{Cakmak2020}
{B. \c{C}akmak,
\textit{Ergotropy from coherences in an open quantum system},
\href{https://doi.org/10.1103/PhysRevE.102.042111}{Phys. Rev. E \textbf{102}, 042111 (2020)}.}


\bibitem{Frenzel2014}
M. F. Frenzel, D. Jennings, and T. Rudolph, \textit{Reexamination of pure qubit work extraction}, \href{http://dx.doi.org/10.1103/PhysRevE.90.052136}{Phys. Rev. E \textbf{90}, 052136 (2014)}.
\bibitem{Olver2010}
F. Olver, D. Lozier, R. Boisvert, and C. Clark, \textit{The NIST Handbook of Mathematical Functions} (Cambridge University Press 2010)


\bibitem{Binder2019}
F. Binder, L. A. Correa, C. Gogolin, J. Anders, and G. Adesso, Ed. \textit{Thermodynamics in the Quantum Regime: Fundamental Aspects and New Directions} (Springer 2019).
%
%





\end{thebibliography}
\end{document}